\documentclass{aa}  
\usepackage{graphicx}
\usepackage[colorlinks=true,citecolor=blue]{hyperref}
\usepackage{multirow}

\usepackage{wasysym}

\usepackage{xcolor}

\usepackage{comment}
\usepackage{float}
\usepackage{dblfloatfix}
\usepackage{multicol,multirow}

\usepackage{amsmath}

\usepackage{placeins}

\usepackage{microtype} 
\microtypesetup{protrusion=true,expansion=true}

\usepackage{txfonts}

\defcitealias{Valles-Perez_2025_accr-i}{Paper I}

\begin{document} 
   \title{The youth of the intracluster medium}
   \subtitle{I. A non-parametric characterisation of the gas and electron number density profiles of $z \simeq 2$ protoclusters}

   \titlerunning{Non-parametric characterisation of proto-ICM density profiles}
   \author{David Vallés-Pérez
          \inst{1,2}\fnmsep\thanks{\email{david.vallesperez@unibo.it}.}
          \and
          Annalisa Bonafede\inst{1,2}
          \and 
          Klaus Dolag\inst{3,4}
          \and
          Marco Balboni\inst{1,5}
          \and 
          Paolo Tozzi\inst{6}
          \and
          Marika Lepore\inst{7,6}
          }

   \institute{Dipartimento di Fisica e Astronomia, Università di Bologna, Via Piero Gobetti 93/2, IT-40129 Bologna, Italy
              \and
              Istituto di Radioastronomia, INAF, Via Piero Gobetti 101, IT-40129 Bologna, Italy
              \and
              Universitäts-Sternwarte, Fakultät für Physik, Ludwig-Maximilians-Universität München, Scheinerstr.1, 81679 München, Germany 
              \and
              Max-Planck-Institut für Astrophysik, Karl-Schwarzschild-Straße 1, 85741 Garching, Germany
              \and INAF - IASF Milano, via A. Corti 12, 20133 Milano, Italy
              \and INAF-Osservatorio Astrofisico di Arcetri, Largo Enrico Fermi 5, 50125, Firenze, Italy
              \and Université Paris-Saclay, Université Paris Cité, CEA, CNRS, AIM, 91191, Gif-sur-Yvette, France
              }

   \date{\today}

  \abstract
   {Protoclusters of galaxies are the earliest phase in the assembly of galaxy clusters, and they can provide invaluable information about plasma physics, cosmic magnetism, and cosmology. However, due to small angular sizes and cosmological dimming, observing the proto-intracluster medium (proto-ICM) associated with protocluster cores is far from trivial.}
   {We aim to provide a non-parametric description of the gas mass and electron number density profiles of the proto-ICM at $z = 2$ and to study their dependence on mass, dynamical state, and central activity.}
   {We extracted and analysed over $3800$ regions around protocluster cores with spherical overdensity masses above $M_\mathrm{500c} > 10^{13} \, M_\odot$ out of a large simulated volume within the \textsc{Magneticum} suite. We studied their density profiles, temperature structure, ionisation degree, and electron number density as a function of mass and other secondary properties characterising dynamical state and central activity extending from the central halo to the surrounding protocluster environment.}
   {Protoclusters present moderate deviations from self-similarity in their density profiles and temperature structure, with a strong double-$\beta$ structure especially relevant at high masses and intense accretion onto the central active galactic nuclei. Hot ionised gas is only dominant at intermediate radii ($r \gtrsim [0.1-0.5] R_\mathrm{500c}$), where its density also correlates with mass and dynamical disturbance.}
   {These results constitute the basis for a forthcoming parametric calibration of proto-ICM density profiles, which could be useful for interpreting observables sensitive to the density and ionisation of the diffuse gas.}

   \keywords{galaxies: clusters: general -- galaxies: clusters: intracluster medium -- galaxies:groups:general -- large-scale structure of Universe -- methods: numerical -- methods: statistical}

   \maketitle

\section{Introduction}
\label{s:intro}

Protoclusters represent the earliest phase in the formation of galaxy clusters. They are commonly identified from large-scale overdensities of galaxies at high redshift ($1.5 \lesssim z \lesssim 8$; e.g. \citealp{Steidel_1998, Chiang_2015, Morishita_2023}) or other tracers biased towards dense environments (e.g. radio galaxies; \citealp{Venemans_2007, Wylezalek_2013}). Such overdensities extend for $\sim (5-30) \, \mathrm{cMpc}$ and are expected to evolve into massive galaxy clusters by $z \approx 0$ (in a statistical sense; see \citealp{Chiang_2013}). At later stages, as massive haloes collapse within them and approach virialisation, their diffuse baryonic component is heated, ionised, and assembled into the nascent intracluster medium (ICM; \citealp{Sunyaev_1972, Chiang_2020}).

There is not a single largely agreed upon definition of protoclusters (see e.g. \citealp{Overzier_2016} for a review). Conceptually, protoclusters may be defined as the Lagrangian regions that will eventually collapse into a massive virialised cluster by $z \sim 0$, and they are typically composed of many small haloes and a large amount of diffuse surrounding material that still has to merge with them. However, this definition is observationally impractical, as the fate of such observed structures can, at best, be constrained only on a statistical basis \citep{Chiang_2013, Muldrew_2015}. Alternatively, protoclusters are sometimes identified with massive quasi-virialised structures at high $z$ \citep{Oteo_2018, Miller_2018}, which are often embedded within larger-scale overdensities. Although $N$-Body simulations show that these high-$z$ haloes do not necessarily evolve into the most massive clusters at $z \sim 0$ \citep{Angulo_2012, Onions_2025}, they more directly correspond to the sites where gas is being assembled and shock-heated, making this definition particularly well suited for studies of the diffuse baryonic component. The term `protocluster' is thus used in the literature to denote these three distinct concepts. Following \citet{Remus_2023}, we refer to large-scale galaxy overdensities as `observed protoclusters', to the corresponding Lagrangian volumes as `protocluster regions', and to the quasi-virialised high-redshift haloes as `protocluster cores' or simply `protoclusters'.

Regarding their baryonic content, it is precisely at the protocluster stage when it is expected to first be shock-heated to temperatures of ${\sim 10^7 \, \mathrm{K}}$ (e.g. \citealp{Kravtsov_2012} for a review). Direct X-ray observations of this proto-ICM are inherently difficult due to a combination of reasons, including small angular sizes, low surface brightness, and foregrounds \citep[e.g.][]{Sarazin_1988}. The thermal \citeauthor{Sunyaev_1972_effect} (\citeyear{Sunyaev_1972_effect}; SZ) effect can also constrain the thermodynamic state of the proto-ICM, but its detection at low masses is likewise challenging \citep{Gobat_2019}. 

Despite these challenges, measurements of hot intracluster gas have recently reached $z \sim (1.5-2)$ \citep[e.g.][]{Gobat_2011, Mantz_2020}. 
Recent works have reported detections of the SZ (at $z \simeq 4$; \citealp{Zhou_2026}) and X-ray (at $z\simeq3$ and $z \simeq 6$, respectively, by \citealp{Travascio_2025} and \citealp{Bogdan_2026}) signals from proto-ICMs.
To date, however, the best-studied proto-ICM at $z \gtrsim 2$ is that of the Spiderweb protocluster. Deep Chandra observations \citep[][$700 \, \mathrm{ks}$]{Tozzi_2022} constrained the average electron number density in its central region ($r < 100 \, \mathrm{kpc}$), while \citet{DiMascolo_2023} confirmed the presence of hot gas by detecting the SZ signal from the same region. Combining both probes, \citet{Lepore_2024} constrained the density, entropy, and pressure profiles out to $r \lesssim 0.6 R_\mathrm{500c}$.\footnote{$R_{\Delta_c}$ is the radius enclosing a total overdensity $\Delta_c$ with respect to the critical density for a flat Lambda-cold dark matter universe.}

As the thermal content of the proto-ICM is assembled, non-thermal components (notably, turbulent motions and magnetic fields) are expected to develop concurrently, analogously to lower-redshift clusters \citep{Bonafede_2010, Govoni_2017, Stuardi_2021}. In this direction, high magnetic fields have been observed in systems approaching $z \sim 1$ \citep{DiGennaro_2021, DiGennaro_2025}. A recent study on the Spiderweb system derived an upper limit of $\mathcal{O}(1 \, \mathrm{\mu G})$ for its proto-ICM \citep{Anderson_2022}. Complementarily, \citet{Pagliotta_2026} derived a lower limit of $0.4 \, \mathrm{\mu G}$ for the central region of CARLA J1510+5958 at $z = 1.72$, suggesting that significant magnetisation is already present in the proto-ICM. However, interpreting these non-thermal components depends critically on the thermodynamic state of the gas. In particular, its density and ionisation structure directly determine the strength of the Faraday rotation effect, and are thus crucial for deriving constraints on protocluster magnetic fields.

The current and foreseeable scarcity of X-ray data together with intrinsic limitations at high redshift (small angular sizes, redshifting of $\sim \mathrm{keV}$ emission out of the observable band) make simulations the only viable way for estimating the density and ionisation state of the proto-ICM. Over the past decades, simulations have been used to explore the definition of protoclusters \citep{Muldrew_2015}, to make comparisons with observations and predict the emergence of a hot gas phase \citep{Saro_2009, Remus_2023}, and to study the galaxy populations of protoclusters \citep{Esposito_2025}. Simulations have also been employed to produce prospects for their detection by, for example, Euclid \citep{Bohringer_2025}, and have highlighted tensions with observations, such as the inability to reproduce the very high star formation rates in some observed systems \citep{Bassini_2020}.

In this first paper of a series, we derive gas mass density, temperature, and electron number density profiles from a sample of $3818$ protoclusters at $z \simeq 2$ from a large-volume simulation of the \textsc{Magneticum} suite \citep{Dolag_2025}. We leverage the statistical power of this sample to characterise the presence of a hot and ionised atmosphere of protoclusters and to describe the departures from self-similarity with mass and secondary parameters describing their larger-scale assembly state and central activity.

The manuscript is structured as follows. In Sect.~\ref{s:methods} we describe the simulation, the protocluster sample, and our post-processing. 
Sects. \ref{s:results} and \ref{s:discussion} then present our main results, and we discuss further aspects, including comparison to observations.
Finally, in Sect.~\ref{s:conclusions} we present a brief summary of the main conclusions.

\section{Methods}
\label{s:methods}

Our analyses are based on the \textsc{Magneticum} simulation suite, whose main features are summarised in Sect.~\ref{s:methods.simulation}, from which we have extracted a sample of protoclusters at $z \simeq 2$ (Sect.~\ref{s:methods.sample}). In Sects. \ref{s:methods.postprocessing} and \ref{s:methods.profile_making}, we discuss our treatment of the baryonic component to compute electron densities, and the technique used to construct radial profiles and stack them. Finally, Sect.~\ref{s:methods.other_properties} covers the computation of other, secondary properties that we used to further constrain the density profiles. 

\subsection{The \textsc{Magneticum Box2b/hr} simulation}
\label{s:methods.simulation}

The \textsc{Magneticum Pathfinder} \citep{Dolag_2016,Dolag_2025} is a large suite of cosmological hydrodynamical simulations designed to model the evolution of galaxies, groups and clusters across a broad range of scales and resolutions, with a special focus on the hot gas content of clusters. In the context of galaxy clusters, they have proved successful in reproducing their observed X-ray \citep[e.g.][]{Biffi_2013, Bahar_2024} and Sunyaev-Zeldovich \citep{Dolag_2016} properties, as well as their metal content \citep{Dolag_2017, Biffi_2025}, among others.

For this study, we focused on \textsc{Box2b/hr} \citep{Remus_2023}, which tracks the evolution of a cosmological volume of $(640 \, h^{-1} \, \mathrm{Mpc})^3 \approx (0.91 \, \mathrm{Gpc})^3$ with $2 \times2880^3$ resolution elements for gas and dark matter (DM), amounting for baryonic and DM mass resolutions of $1.4 \times 10^8 M_\odot$ and $6.9 \times 10^8 M_\odot$. As in the rest of the \textsc{Magneticum Pathfinder} suite, a WMAP7 cosmology \citep{Komatsu_2011} is assumed, with matter and dark energy density parameters $\Omega_m=0.272$ and $\Omega_\Lambda=1-\Omega_m$, cosmic baryon fraction $f_b = 0.168$, Hubble dimensionless constant $h=0.704$ ($H_0 = 100 h \, \mathrm{km \, s^{-1} \, Mpc^{-1}}$), and initial conditions stemming from a primordial power spectrum with $\sigma_8 = 0.809$ and $n_s=0.963$. 

The simulation was evolved with a modified version of \textsc{Gadget} \citep{Springel_2005_gadget}, including a modern, low-viscosity formulation of the smoothed particle hydrodynamics (SPH) method \citep{Dehnen_2012, Beck_2015}, detailed chemical enrichment following \citet{Tornatore_2007}, cooling \citep{Wiersma_2009} and a multiphase model for star formation and feedback from supernovae \citep{Springel_2003}, as well as supermassive black holes modelled as sink particles releasing thermal active galactic nucleus (AGN) feedback \citep{Springel_2005_agn, DiMatteo_2005, Fabjan_2010}. We refer the reader to \citet{Dolag_2025} for an extensive review on the properties of the simulation model, together with its validation against local and global, galaxy and galaxy cluster scaling relations through cosmic time.

\subsection{Protocluster sample selection and halo catalogues}
\label{s:methods.sample}

We selected protoclusters as the structures around the density peaks of high-$z$, virialised\footnote{Our definition of virialised structures is based on a density contrast under the spherical symmetry assumption \citep{Lacey_1993, Bryan_1998} customarily made in simulations. Note, however, that these structures, especially at high redshift, are not closed systems and may lie at different dynamical states (i.e. not necessarily in virial equilibrium). Similarly, our definition of protoclusters only comprises a small fraction of the total amount of matter that will reside in virialised haloes at $z \simeq 0$ \citep[e.g.][]{Muldrew_2015}. For a more thorough discussion on these issues, we refer the reader to \citet{Remus_2023}.} DM haloes (roughly corresponding to the definition of protocluster cores by \citealp{Remus_2023} and their immediate surroundings), without imposing any criterion on their larger-scale environment. Throughout this manuscript, we focus on the objects at $z \simeq 2$,\footnote{In particular, we extract the catalogues from snapshot 11 of \textsc{Box2b/hr}, at $z = 1.98$.} while in a future work we plan to extend these results to a broader redshift interval. From the whole \textsc{Box2b/hr}, we selected all regions around \textsc{Subfind} \citep{Springel_2001, Dolag_2009} friends-of-friends groups with mass $M_\mathrm{500c} > 10^{13} \, h^{-1} \, M_\odot$, which we identify as our sample candidates. 

These correspond to a total of $3836$ regions, which were extracted out to a radius of $8 R_\mathrm{500c}$ from the \textsc{Subfind} centre. In a small fraction of cases ($\lesssim 5\%$), due to strongly irregular mass distributions, \textsc{Subfind} centres depart from the apparent centre of spherical symmetry of the halo, and it is biased instead towards nearby, prominent substructures (see App. \ref{s:app.matching_asohf} for an example and further discussion). This is why we reprocessed all regions with the publicly available spherical overdensity halo finder \textsc{ASOHF}\footnote{\url{https://github.com/dvallesp/ASOHF}.} \citep{Planelles_2010, Valles-Perez_2022}, which locates the halo centre from the global density peak (i.e. defined at large scales). Incidentally, \textsc{ASOHF} also provides a complete characterisation of substructures and galaxies, which is exploited in Sect.~\ref{s:methods.other_properties}. Out of the $3836$ regions, in $18$ occasions there was either no clear one-to-one match between \textsc{ASOHF} haloes and \textsc{Subfind} original positions (15 cases), due to strongly irregular morphologies, merging systems, or overlapping \textsc{Subfind} regions, or at least one of the properties in Sect.~\ref{s:methods.other_properties} could not be properly computed (three cases). These instances were removed from the catalogue, leaving a sample of $3818$ objects. 

\subsection{Gas treatment and computation of electron densities}
\label{s:methods.postprocessing}

The \textsc{Magneticum Pathfinder} simulations implement a multiphase, multispecies modelling of the cosmic plasma. Regarding its multiphase aspect, following \citet{Springel_2003}, high-density SPH particles can develop a subresolution cold phase (which in turn fuels star formation) out of the diffuse phase. To account for this two-phase nature of the gas in our post-processing, given a gas particle with cold mass fraction ${f_\mathrm{c} = m_\mathrm{c} / m_\mathrm{t}}$, mass-weighted temperature $T_\mathrm{mw}$, and density $\rho_t$, whenever we aim to retrieve the properties of diffuse gas, we treated it as a particle with diffuse gas mass, temperature, and densities given by
\begin{equation}
    m_\mathrm{h} = (1 - f_\mathrm{c})m_\mathrm{t}, \qquad T_\mathrm{h} = \frac{T_\mathrm{mw}}{1 - f_\mathrm{c}}, \qquad \rho_\mathrm{h} = (1 - f_\mathrm{c}) \rho_\mathrm{t},
    \label{eq:hotgas}
\end{equation}
\noindent that is, we assumed the cold ISM has a negligible temperature and volume fraction.

The computation of electron densities, $n_e$, is not straightforward once the assumptions of full ionisation and primordial composition are dropped. In particular, the chemical enrichment model in \textsc{Magneticum} tracks eleven elemental abundances (H, He, C, N, O, Ne, Mg, Si, S, Ca, and Fe). 
In post-processing, to determine the relative ionisation states of each element, we assumed collisional ionisation equilibrium (CIE) and solved the equations for the ionisation balance on a grid of temperatures, separately for each chemical species, to obtain the quotient $(n_e / \rho)_\text{X} = f(T)$ for each tracked element $X$. We computed these interpolation tables for the interval $10^4 \leq T/\mathrm{K} \leq 10^9$. Below the lower limit, we used $n_e / \rho = 0$, while above $10^9 \, \mathrm{K}$ the elements are considered to be fully ionised. This only introduces noticeable errors in the case of iron (where they are still largely below $1 \%$). The total electron density of each SPH particle was thus computed from these interpolation tables, taking into account its diffuse gas temperature, $T_h$, and its detailed chemical composition (which is, in turn, a result of the stellar assembly history of the protocluster and its members).

The calculation therefore relies on two key assumptions, namely, (i) neglecting photoionisation effects and (ii) the assumption of CIE. Their validity is discussed in greater detail in Sect.~\ref{s:discussion.assumptions}.

\subsection{Profile making and stacking}
\label{s:methods.profile_making}

For each protocluster in the sample, we extracted three-dimensional radial profiles of DM, stellar, gas and total mass densities, diffuse gas density (using several temperature thresholds), and electron number density. The profiles are computed from spherical shells centred on the DM density peak, and the width of the shells is selected so that each radial bin encompasses at least ${N_\mathrm{min\,part}^\mathrm{bin} = 1000}$ gas particles, and is at least $\Delta \log_{10} r = 0.01 \, \mathrm{dex}$ wide. Within the $i$-th bin, $r_\mathrm{in,i} \leq r < r_\mathrm{out,i}$, its density is computed as $\rho_i = \sum_\alpha m_\alpha / \left[ 4 \pi / 3 \left( r_\mathrm{out,i}^3 - r_\mathrm{in,i}^3 \right) \right]$, and it is assigned to the centre-of-mass radius of the bin, $r_i = \sum_\alpha m_\alpha r_\alpha / \sum_\alpha m_\alpha$, where $m_\alpha$ and $r_\alpha$ are the mass and the radial position of the $\alpha$-th particle.

In order to stack the profiles, each individual profile was resampled by linear interpolation in double-logarithmic space to a common radial grid with $r_\mathrm{min} = 0.01 R_\mathrm{500c}$, $r_\mathrm{max} = 8 R_\mathrm{500c}$, and $\Delta \log_{10} r = 0.01 \, \mathrm{dex}$. The values at each rescaled $r/R_\mathrm{500c}$ were obtained, as in \citet{Valles-Perez_2026}, by computing the robust mean \citep{Beers_1990} of the corresponding values of $\log_{10} \rho$. The choice of a large $r_\mathrm{max}$ allowed us to probe not only the central halo but also the surrounding environment.

\subsection{Other protocluster properties}
\label{s:methods.other_properties}

To characterise the secondary dependences of the profiles in properties other than the mass of the protocluster, we determined additional quantities associated to the global assembly state, and the inner state of the ICM. We briefly discuss them below.

\paragraph{Mass ratio, $M_{12}$.} At low redshift, the ratio between the masses of the brightest cluster galaxy (BCG) and the second most massive galaxy (i.e. the most massive satellite) has often been invoked as an indicator of recent mergers (insofar the most massive satellite is often the BCG of a recently accreted group; \citealp{Jones_2003, Hearin_2013, Ragagnin_2019}). Since mergers have been shown to imprint important features on the density and thermodynamic profiles of the low-$z$ ICM \citep{Lau_2015, Valles-Perez_2026, Correa-Magnus_2026}, we also chose this variable to quantify the larger-scale assembly state of protoclusters. In App. \ref{s:app.M12} we briefly discuss its correlations with other standard dynamical state indicators.

Operationally, we determine $M_{12}$ from the \textsc{ASOHF} galaxy catalogues. 
The BCG is selected as the centrally located, most massive galaxy\footnote{Both requirements are combined by choosing the galaxy minimising the quotient $r_{\mathrm{gal},i} / M_{\mathrm{gal},i}$.}
and, unlike other halo finders, in \textsc{ASOHF} its mass does not contain the diffuse component filling the whole halo (i.e. the intracluster light). Instead, \textsc{ASOHF} determines an outer boundary for central galaxies that considers the most conservative of several thresholds (a minimum stellar overdensity of $1000$ with respect to the cosmic background mass density, maximum physical radius of $200 \, \mathrm{kpc}$, or an increase in the radial profile of stellar mass density indicating the presence of a satellite). Any galaxy within $R_\mathrm{200c}$ of the BCG centre is considered to be a satellite. The mass ratio $M_{12}$ is hence defined as the quotient between the bound stellar masses of most massive satellite, and that of the BCG, $M_{12} \equiv M_{*,\text{1st sat}} / M_{*,BCG}$. To prevent $M_{12}$ from being unbounded, it is capped at a value of $M_{12} = 1$ in the few ($\sim 1\%$) occasions where $M_{*,\text{1st sat}} > M_\mathrm{BCG}$.

\paragraph{Integrated Eddington ratio, $\bar f_\mathrm{Edd}$.} The core regions of local clusters are highly affected by the heating from AGNs (e.g. \citealp{Sijacki_2007}). While most current observations of the proto-ICM come precisely from the vicinity of radio-loud galaxies \citep[e.g.][]{Wylezalek_2013, Hatch_2014, Chapman_2025}, the effect of AGN feedback (or lack thereof) on protocluster cores has not been fully explored due to the inherent observational challenges. To characterise the impact of AGN feedback on the profiles, we use the Eddington mass ratio, defined here as the quotient between the supermassive black hole (SMBH) accretion rate $\dot M_\mathrm{BH}$, and its value at the Eddington limit, $\dot M_\mathrm{BH}^\mathrm{Edd} = 4 \pi G m_p M_\mathrm{BH} / (\epsilon_r c \sigma_T)$, where $M_\mathrm{BH}$ is the mass of the SMBH, $\epsilon_r = 0.2$ is its radiative efficiency (matching the one used in the simulation), and $c$, $\sigma_T$, $G$ and $m_p$ respectively are the speed of light, Thomson electron scatter cross-section, gravitational constant, and proton mass. To ameliorate the scatter induced by the stochasticity of BH accretion on short timescales, we determined an integrated (average) Eddington ratio in the following way:

\begin{enumerate}
    \item We selected the central SMBH of each protocluster as the most massive and closest to the DM density peak black hole particle (using the same criterion as for the BCG above).\footnote{With this definition, the central SMBH could not coincide with that of the BCG. This occurs in $\lesssim 5\%$ of the cases, typically in systems with multiple massive galaxies, where the BCG is not uniquely defined and may be offset from the halo centre, while the SMBH remains central.} This yields a SMBH mass
    $M_{\mathrm{BH},2}$ at $t=t_2$ (i.e. the snapshot being analysed).
    \item We located the same SMBH particle in the immediately previous snapshot (around $\Delta t \sim 500 \, \mathrm{Myr}$ before), yielding a mass $M_{\mathrm{BH},1}$ at $t=t_1$. In the event that the SMBH particle was seeded in between the two snapshots considered, we assign $M_{\mathrm{BH},1}$ as the seeding mass ($2 \times 10^5 \, h^{-1}\, M_\odot$), and $t_1$ to the cosmic time when the particle was spawned.
    \item We estimated the average Eddington ratio ${f_\mathrm{Edd} \propto \dot M_{\mathrm{BH}} / M_{\mathrm{BH}}}$ by assuming it to be constant for $t_1 < t < t_2$ as
    \begin{equation}
        \bar f_\mathrm{Edd} = \frac{\epsilon_r c \sigma_T}{4\pi G m_p} \frac{\ln \left( M_{\mathrm{BH},2} / M_{\mathrm{BH},1} \right)}{t_2-t_1},
    \end{equation}
    \noindent with $\ln$ representing the natural (base-$e$) logarithm.
\end{enumerate}

\section{Results}
\label{s:results}

We begin this section with a discussion on the general trends of the gas mass density and temperature profiles with halo mass in Sect.~\ref{s:results.density_mass} that justify the behaviours reported later in Sect.~\ref{s:results.ne_mass} for the electron number density profiles. In Sect.~\ref{s:results.other_dependencies} we focus instead on the secondary dependences of the proto-ICM profiles on assembly state and central activity.

\subsection{Density and temperature structure}
\label{s:results.density_mass}

\begin{figure*}
    \centering
    \includegraphics[width=0.33\linewidth]{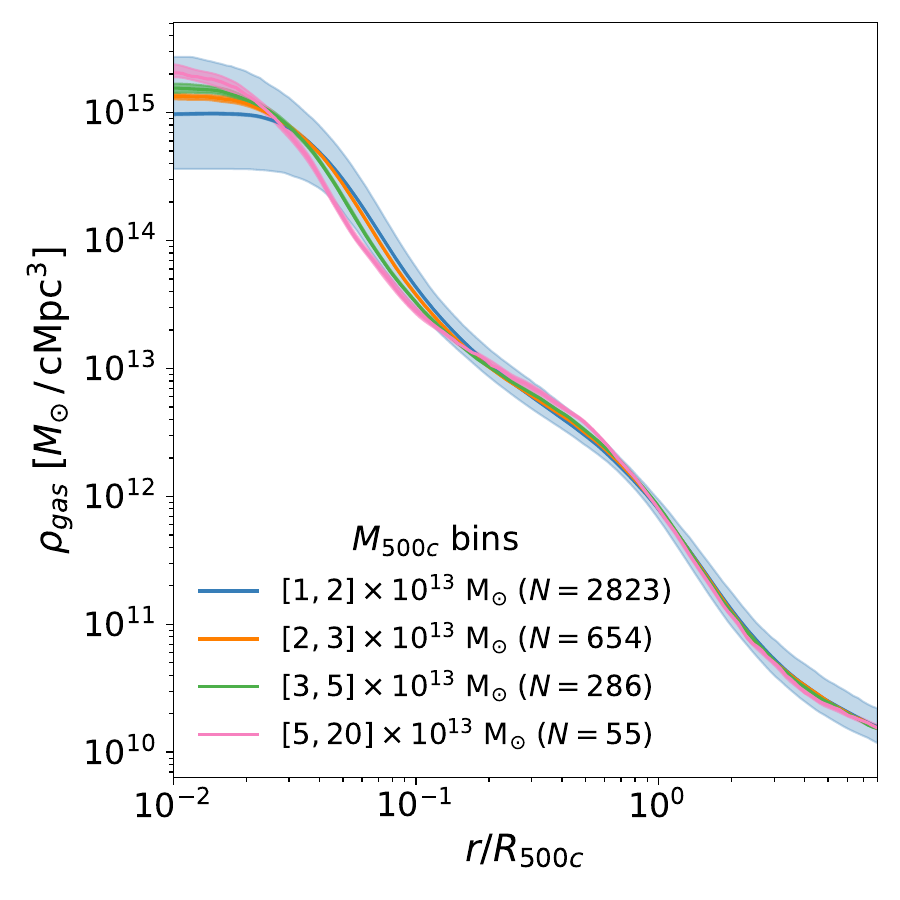}~
    \includegraphics[width=0.33\linewidth]{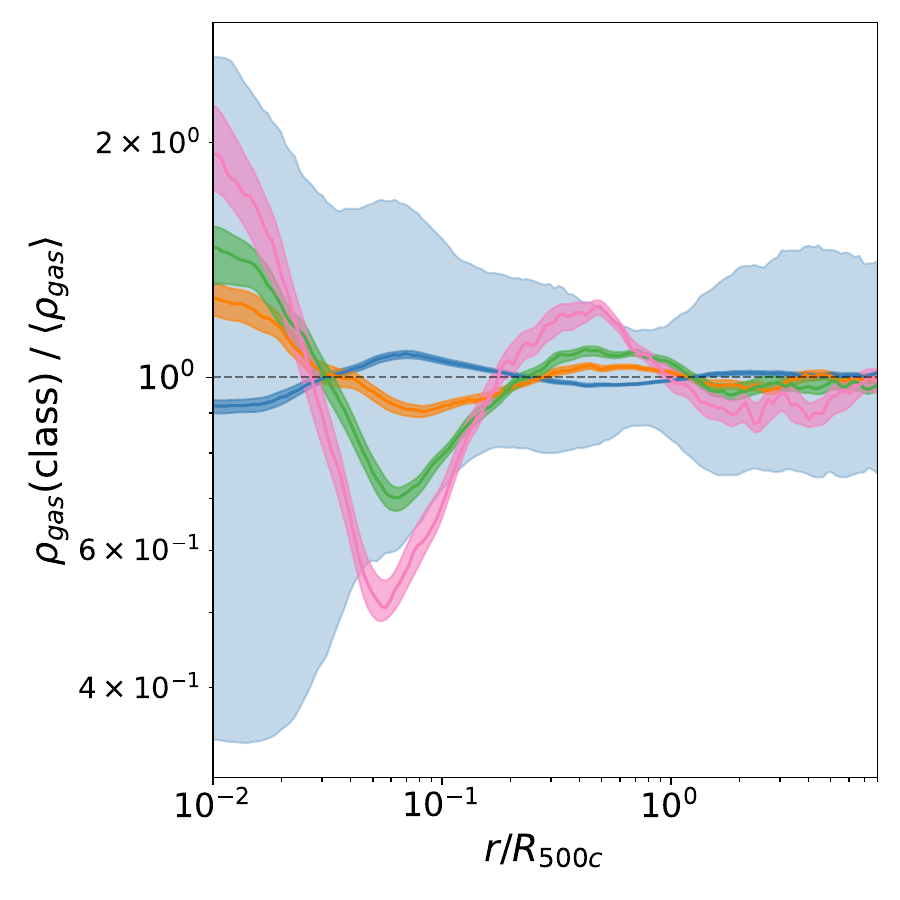}~
    \includegraphics[width=0.33\linewidth]{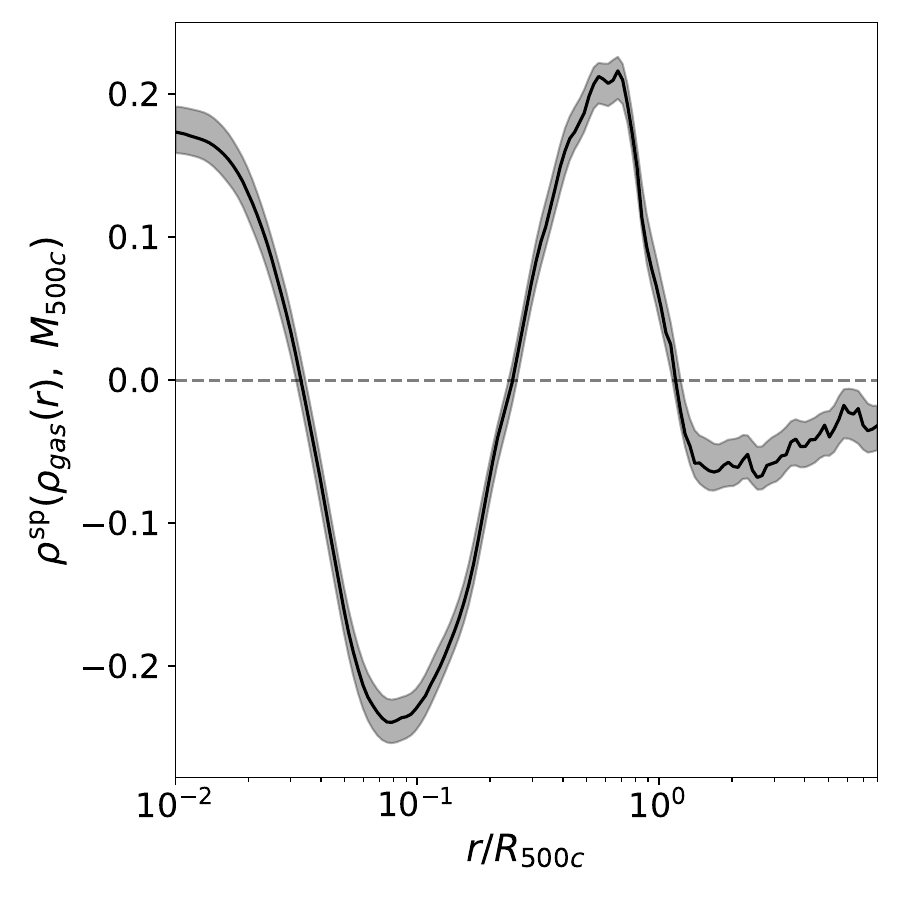}
    \caption{Mass dependence of the protocluster gas density profiles. 
    \textit{Left:} Comoving gas density profiles, stacked by protocluster mass $M_\mathrm{500c}$. The legend also indicates the number of objects per mass bin. Dark shades indicate the uncertainty in the stacked profiles, while light shaded regions reflect the population scatter. 
    \textit{Centre:} Same as the left panel but all profiles have been normalised by the stack over the whole sample.
    \textit{Right:} Spearman correlation between the gas densities at each $r$ and mass computed over the whole sample, signaling the departures from self-similarity.}
    \label{fig:gas_density}
\end{figure*}

We first examine the behaviour of comoving gas densities\footnote{
Although the physically meaningful quantity is the physical density, $\rho_\mathrm{phys}=\rho_\mathrm{com} (1+z)^3$, unless otherwise specified, here we always normalise densities by the comoving volume. This is a conventional choice regarding the presentation of the results and bears no impact on subsequent results, as we work at a single redshift, $z = 1.98$. 

Expressing densities in comoving units also facilitates comparison across redshifts, as the dominant evolution of the ICM density (especially within $r \lesssim 2 R_{200c}$) is approximately captured by the self-similar $(1+z)^3$ evolution \citep[see e.g.][]{Lau_2015, Mostoghiu_2019}.
} around our protoclusters as a function of their total mass ($M_\mathrm{500c}$) in Fig.~\ref{fig:gas_density}. In the left panel, different lines correspond to bins in $M_\mathrm{500c}$,\footnote{The catalogue is selected using a threshold on the Subfind mass, $M_\mathrm{500c}^\mathrm{Subfind} > 10^{13} \, h^{-1} M_\odot$, whereas all masses quoted elsewhere in the manuscript are expressed in $M_\odot$. Therefore, although the nominal $[1, \, 2] \times 10^{13} \, M_\odot$ bin corresponds to objects selected above $M_\mathrm{500c}^\mathrm{Subfind} > 10^{13} \, h^{-1} M_\odot \approx 1.4 \times 10^{13} \, M_\odot$, their actual $M_\mathrm{500c}^\mathrm{ASOHF}$ used for the binning can fall below this threshold, still remaining always above $M_\mathrm{500c} > 10^{13} M_\odot$.} where the dark shaded regions represent the uncertainty in the stacked profiles estimated by bootstrap resampling. For one of the classes, we represent the population scatter as the lighter shaded region. Overall, the profiles show a clear double-$\beta$ structure, where the external component has a characteristic radius $\sim (0.5-1)R_\mathrm{500c}$, while the central steepening occurs at $r \lesssim 0.1 R_\mathrm{500c}$. Such a feature is also observed in some relaxed cool-core systems at low redshift \citep[e.g.][]{Vikhlinin_2006} but seems to be more frequent in our simulated systems at $z \simeq 2$. When comparing these results, it is important to note that observed $\rho_\mathrm{gas}(r)$ profiles (e.g. \citealp{Vikhlinin_2006, Pratt_2022}) are typically inferred assuming $\rho_\mathrm{gas} \propto n_e$ and therefore only trace the hot ionised gas. In contrast, our $\rho_\mathrm{gas}$ includes all gas phases, including those outside the X-ray-emitting phase. This is why we defer any comparison to data from observations to Sect.~\ref{s:results.ne_mass}.

To complement this view, the central panel shows the stacked profiles, normalised by the full-sample stack. The gas density profiles deviate from self-similarity, with higher-mass protoclusters exhibiting cuspier and overall higher inner gas densities (by a factor of $\sim 2$ compared to their lower-mass counterparts). This behaviour is driven by a stronger steepening around $r/R_\mathrm{500c} \sim 0.1$ in more massive protoclusters (where densities are instead a factor of $\sim 2$ smaller than in lower-mass systems). At intermediate radii, $(0.2-1) R_\mathrm{500c}$, the highest-mass protoclusters also show an excess gas density, by up to $\sim 25\%$, relative to the lowest-mass ones. These systematic trends, however, only amount to a modest contribution of the overall large scatter around the profiles (Spearman rank correlations up to $| \rho^\mathrm{sp} | \sim 0.2$; see the right panel of Fig.~\ref{fig:gas_density}). Taken together with the non-trivial mass-dependence of the profiles, this points at a combination of several effects, which we analyse in more detail in the following sections.

\begin{figure}
    \centering
    \includegraphics[width=0.78\linewidth]{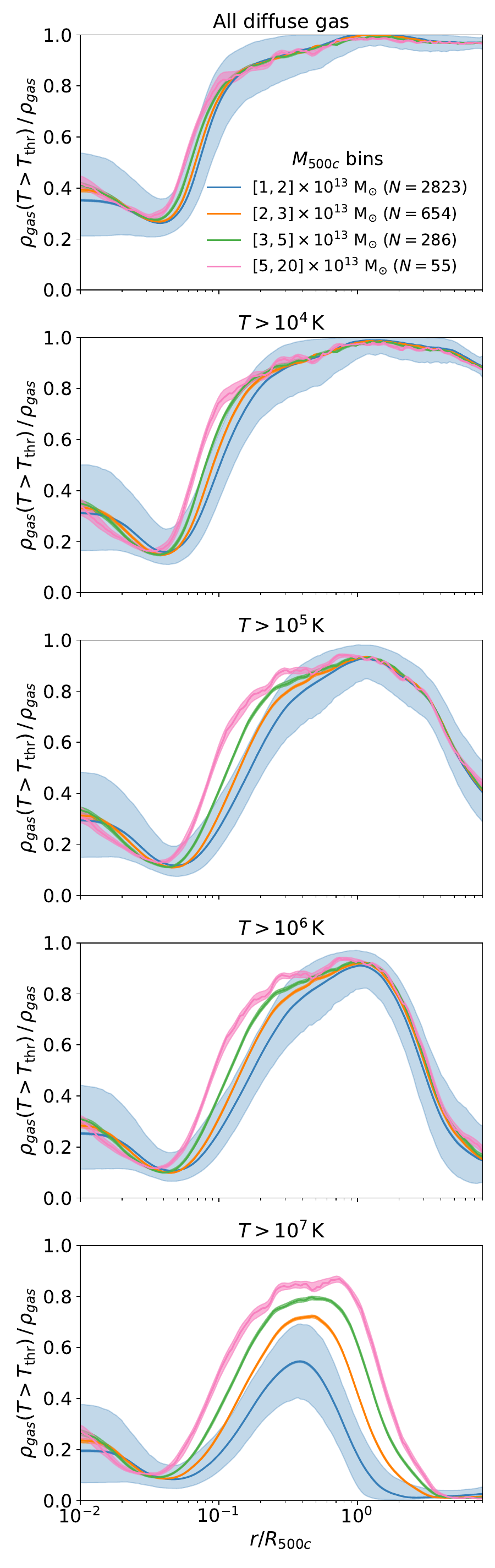}
    \caption{Density fraction above different temperature thresholds stacked by protocluster mass $M_\mathrm{500c}$ according to the legend. \textit{Top panel:} All diffuse gas (except the cold subresolution ICM). \textit{Second panel:} Diffuse gas above $10^4 \, \mathrm{K}$ (all warm and hot gas). \textit{Third panel:} Diffuse gas above $10^5 \, \mathrm{K}$ (all primordial gas is ionised). \textit{Fourth panel:} Diffuse gas above $10^6 \, \mathrm{K}$ (hot, X-ray emitting gas). \textit{Bottom panel:} Diffuse gas above $10^7 \, \mathrm{K}$ (bremsstrahlung-emitting gas).}
    \label{fig:gas_density_fractions_T}
\end{figure}

Besides the overall density structure of the nascent ICM, its temperature structure is crucial to determining its ionisation state and emission properties. In Fig.~\ref{fig:gas_density_fractions_T} we analyse the fraction of the total gas density above several temperature thresholds (from top to bottom): all diffuse gas (excluding the cold, subgrid ISM; see Sect.~\ref{s:methods.postprocessing}), $T>10^4 \, \mathrm{K}$ (all warm and hot gas), $T>10^5 \, \mathrm{K}$ (fully ionised primordial gas), $T>10^6 \, \mathrm{K}$ (roughly, the hot X-ray emitting plasma), and $T>10^7 \, \mathrm{K}$ (where bremsstrahlung dominates over line emission). A complementary study of the mass fractions in different temperature intervals (instead of fractions above given thresholds) is provided in App. \ref{s:app.density_mass_additional.temperature}.

The top panel shows that the central regions of protoclusters ($r \lesssim 0.05 R_\mathrm{500c}$) contain $(60-80)\%$ of their gas in a condensed, cold, and potentially star-forming phase. These fractions drop sharply at ($r /R_\mathrm{500c} \sim 0.05-0.1$), where the proto-ICM becomes dominant, its fraction gradually increasing from $\sim 80\%$ at ${\sim 0.1 R_\mathrm{500c}}$ to $\sim 100\%$ at $R_\mathrm{500c}$ regardless of mass. The fraction of warm/hot gas ($T > 10^4 \, \mathrm{K}$, second panel) presents a minimum at similar radii of the steepening in the total gas density ($r /R_\mathrm{500c} \sim 0.05-0.1$). 
Interestingly, we tested that the location of this feature is not self-similar (i.e. does not scale as $r \propto R_\mathrm{500c}$). Instead, the statistically preferred scaling is consistent with a nearly constant radius $r \sim (70\pm20) \, \mathrm{ckpc}$.\footnote{In particular, we found that, if $r_\mathrm{d,i}$ is the radius of this feature for the $i$-th protocluster, then the scatter in in the values of $\{ R_\mathrm{500c,i}^\alpha r_\mathrm{d,i} \}_{i=1}^{N_\mathrm{PCs}}$ is minimised for $\alpha = 0.02\pm0.01$, whereas a self-similar scaling would imply $\alpha=-1$.} This suggests it is decoupled from the bulk proto-ICM, and likely marks the transition between inner regions (dominated by the BCG, effectively heated by AGN feedback) and the outer parts of the protocluster core (where dense gas cools down efficiently, and has not yet been heated by AGN-driven bubbles).
We explore their connection with AGN feedback in more detail in Sect.~\ref{s:results.other_dependencies}.

The third and fourth panels approximately show the fraction of gas for which H and He are fully ionised ($T \gtrsim 10^5 \, \mathrm{K}$) and hot, X-ray emitting ($T \gtrsim 10^6 \, \mathrm{K}$). The dip at $\sim 70 \, \mathrm{ckpc}$ is naturally also present at these larger temperature thresholds, and its strength anti-correlates weakly with mass, showing a greater defect of ionised and hot gas for lower-mass protoclusters, potentially indicating a stronger effect of AGN feedback. At intermediate radii ($r/R_\mathrm{500c} \in [0.1, 1]$), the fractions also depend strongly on mass. Whereas massive protoclusters ($M_\mathrm{500c} \gtrsim 5 \times 10^{13} \, M_\odot$ at $z=2$) have most ($\sim 80\%$) of their gas in the hot phase already at $0.1 R_\mathrm{500c}$, for our lower masses bins, high fractions of gas above $10^6 \, \mathrm{K}$ are only reached around $\sim 0.5 R_\mathrm{500c}$. Finally, at the temperature threshold for bremsstrahlung to roughly dominate the X-ray emission ($T \gtrsim 10^7 \, \mathrm{K}$, fifth panel), there's a clear dependence of the mass fraction of gas above this threshold and $M_\mathrm{500c}$. Indeed, the virial temperature for a $10^{13} \, M_\odot$ halo within $R_\mathrm{500c}$ at $z = 2$ roughly reaches $10^{7} \, \mathrm{K}$. Therefore, only for the higher mass objects we expect an atmosphere dominated by hot, bremsstrahlung emitting gas, while for protoclusters below $2 \times 10^{13} \, M_\odot$ we only find half of their mass to exceed this temperature threshold at intermediate radii, and thus their emission is likely to be strongly dominated by lines. Complementarily, further details on the mass-dependence of gas density profiles and baryon depletion functions are discussed in App. \ref{s:app.density_mass_additional.density}.

\subsection{Electron number density}
\label{s:results.ne_mass}

\begin{figure*}
    \sidecaption
    \parbox{12cm}{%
        \includegraphics[width=6cm]{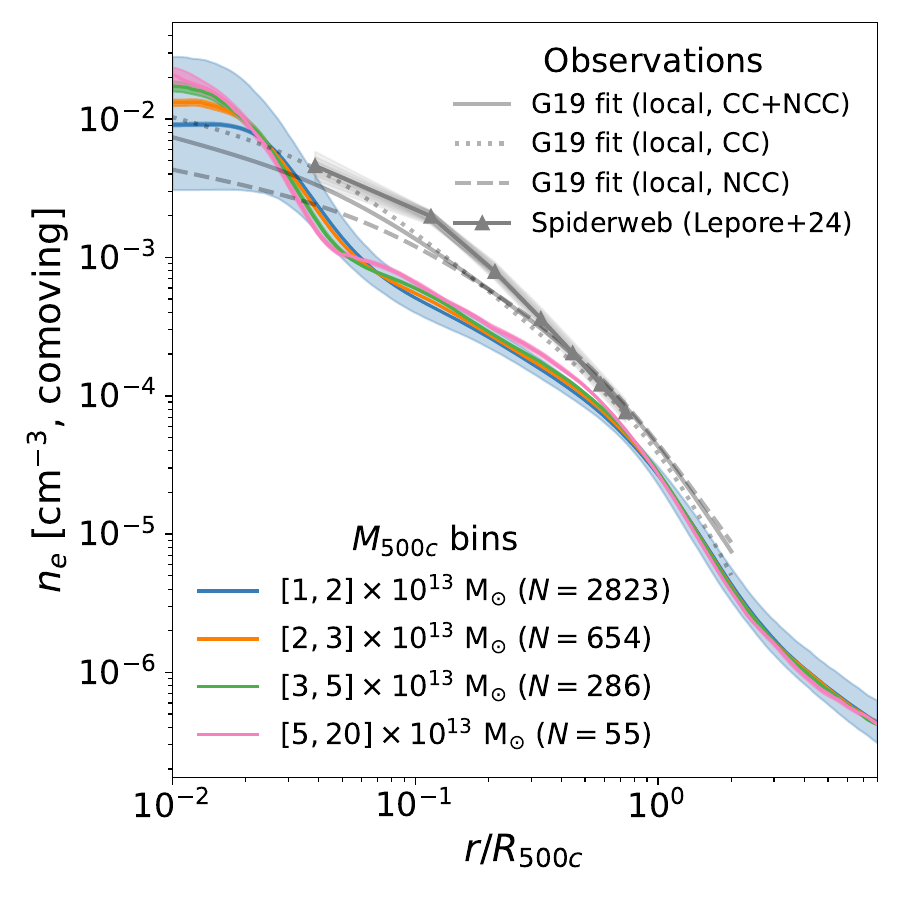}%
        \includegraphics[width=6cm]{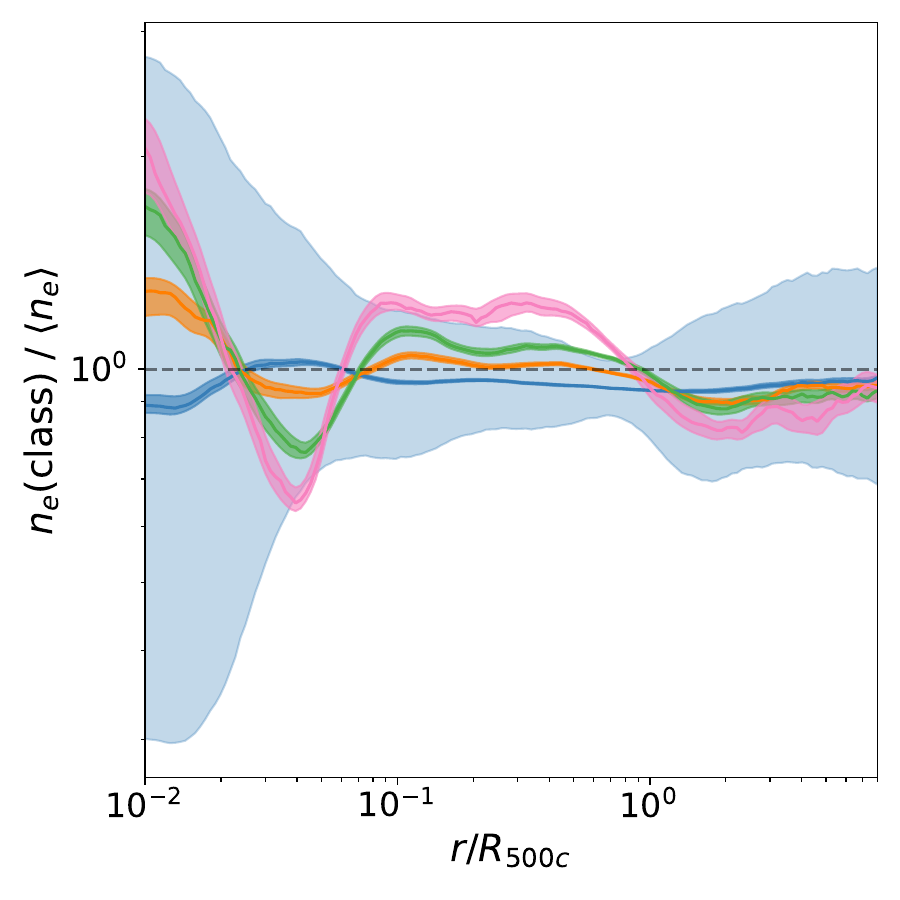}
        \includegraphics[width=6cm]{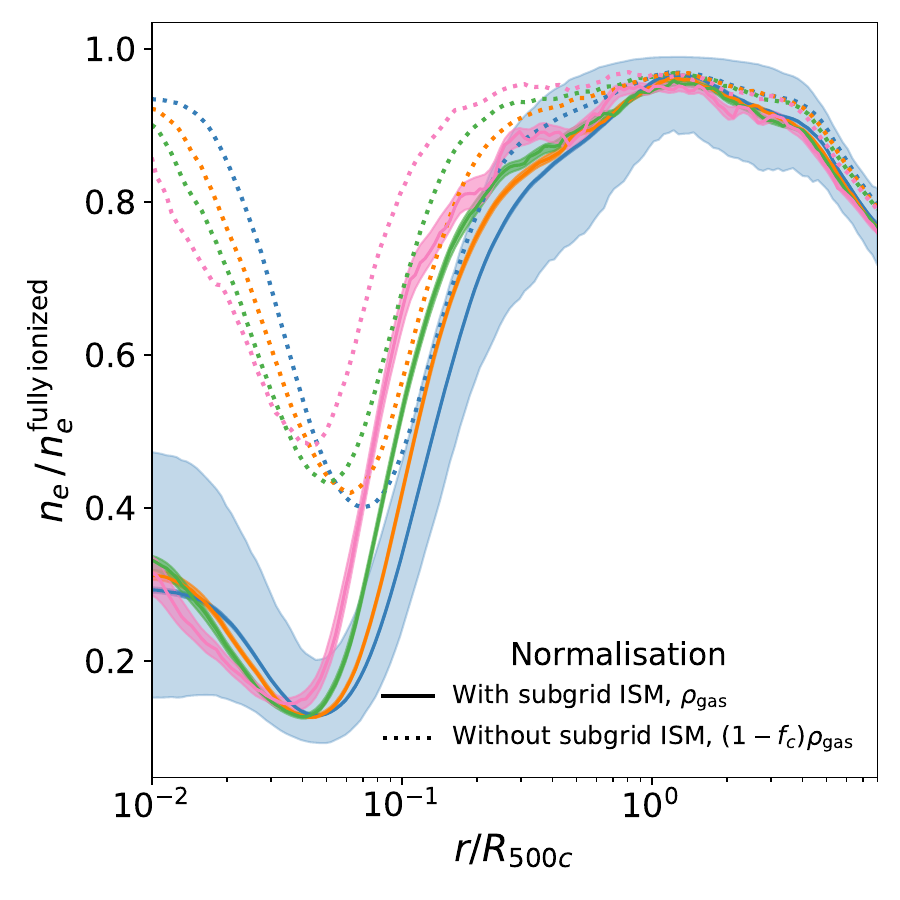}%
        \includegraphics[width=6cm]{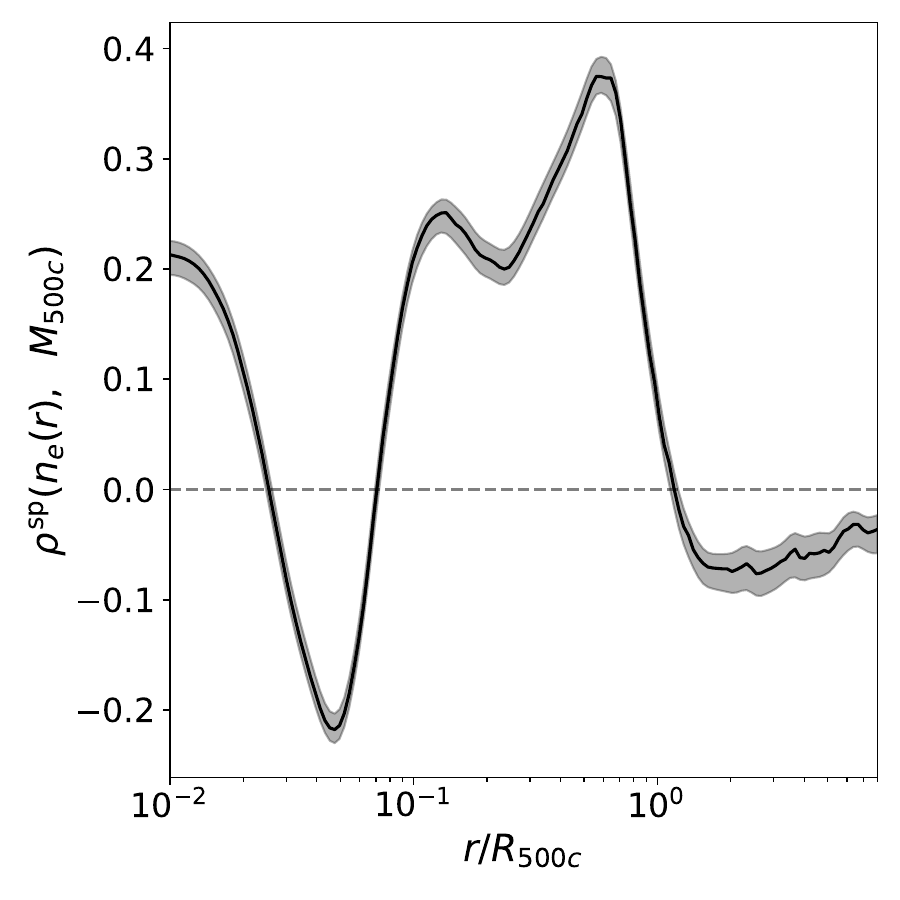}
    }
    \caption{Analysis of the electron number density profiles, $n_e(r)$, in our protocluster sample as a function of protocluster mass $M_\mathrm{500c}$. \textit{Top-left:} Electron comoving number density profiles. Comparisons to observational data of \citet{Ghirardini_2019} and \citet{Lepore_2024} are shown in grey according to the upper-right legend.
    \textit{Top-right:} Same as the top-left panel but all profiles have been normalised by the stack over the whole sample. \textit{Bottom-left:} Ratio of the actual electron densities to ones corresponding to a fully ionised plasma, derived from the total gas density (solid lines) or the diffuse gas density (dotted lines). \textit{Bottom-right:} Spearman correlation between the electron number density profiles at each $r$ and mass.}
    \label{fig:electrondensity_mass}
\end{figure*}

As a consequence of the total gas density and temperature trends described in Sect.~\ref{s:results.density_mass}, we now examine the radial behaviour of electron number densities across the protocluster sample. The top-left panel of Fig.~\ref{fig:electrondensity_mass} shows the $n_e(r)$ profiles in the same mass bins as the preceding analyses, while the top-right panel presents them normalised by global stacked profile, as also done in Fig.~\ref{fig:gas_density}. Typical comoving electron densities range from $(1-2) \times 10^{-2} \, \mathrm{cm}^{-3}$ (physical densities $[0.3-0.6] \, \mathrm{cm}^{-3}$) in the protocluster centre, to $10^{-5} \, \mathrm{cm}^{-3}$ ($3 \times 10^{-4} \, \mathrm{cm}^{-3}$ in physical units) at $r \approx R_\mathrm{500c}$.

Central electron densities significantly increase with $M_\mathrm{500c}$. On average, the most massive systems are over twice as dense as the least massive ones, exhibiting cuspy, strongly double-$\beta$-shaped profiles, while lower-mass objects do not steepen as much towards internal radii and tend towards flat cores. The object-to-object scatter in the centre is remarkably large ($\sim 1 \, \mathrm{dex}$ for the lowest-mass bin; light-blue shaded regions in the top-right panel), leading to a weak, positive Spearman rank correlation of $\sim 0.2$ (bottom-right panel). 

The bottom-left panel, shows the ratio of electron densities to those of a fully ionised plasma, ${n_e^\mathrm{fully \, ionised} = \rho_\mathrm{gas} / (\mu_e m_p)}$, derived either from the total gas density (solid lines) or only the diffuse gas density (dotted lines). While the latter can be interpreted as an ionisation fraction, the former also accounts for the mixing between two co-spatial phases in the centralmost regions: namely, the ISM and the proto-ICM. 

This shows that only $\sim 30\%$ of the total gas is diffuse and ionised in protocluster centres, where radiative cooling is efficient due to the high densities and gas evacuates the diffuse phase into the ISM phase. The remaining diffuse gas, however, is hot and mostly ionised, especially in the least massive systems.
Owing to this, from $\sim 0.1 R_\mathrm{500c}$ to $\sim 0.01 R_\mathrm{500c}$, electron densities increase only by a factor of $\sim 10$, instead of the nearly two orders of magnitude increase of the gas mass density (Fig.~\ref{fig:gas_density}). 

At more intermediate radii, $0.1 \lesssim r/R_\mathrm{500c} \lesssim 1$, more massive protoclusters present electron densities overall larger by $\sim 30\%$ when compared to poorer systems, with correlations ranging $0.3-0.4$. This is contributed by a combination of the effects on the total density structure reviewed in Sect.~\ref{s:results.density_mass}, and larger proto-ICM fractions and ionisation degrees in more massive systems: as shown by the dotted lines in the bottom-left panel of Fig.~\ref{fig:electrondensity_mass}, the ionisation fraction of higher-mass protoclusters reaches $\sim 80\%$ already at $\sim 0.1 R_\mathrm{500c}$, whereas comparable levels of ionisation are generally attained only around $r \sim 0.2 R_\mathrm{500c}$ in the lowest-mass systems. The fractions computed with respect to the $n_e$ derived from the total gas density (solid lines) highlight that, for the lowest-mass protoclusters, the dominance of the ISM pushes the beginning of the region dominated by ionised proto-ICM further out ($r \gtrsim 0.3 R_\mathrm{500c}$). The combination of both these effects (decreasing density, increasing ionisation fraction) makes the $n_e(r)$ profiles of high-mass protoclusters remarkably flat around $\sim 0.1 R_\mathrm{500c}$, a feature which is not found for the mass density profiles.

Therefore, the centre of protoclusters exhibits a strong radial gradient in diffuse and ionised gas fractions, increasing from $\sim 30\%$ in the centre to nearly unity at $(0.1-0.5)R_\mathrm{500c}$, with the minimum ionisation typically occurring at $\sim 70 \, \mathrm{kpc}$, just outside the region directly heated by AGNs.
Correlations of the $n_e(r)$ profile with mass are moderately low outside $\sim R_\mathrm{500c}$, implying a roughly self-similar behaviour. However, the central features do not seem to align with $R_\mathrm{500c}$, and in particular the electron density defect at $\sim 70 \, \mathrm{kpc}$ does not scale with cluster mass.

The solid, dashed and dotted grey lines in the upper left panel show the best-fitting universal profiles by \citet{Ghirardini_2019} for, respectively, the complete, the non-cool-core (NCC) and the cool-core (CC) samples of local ($z < 0.1$) clusters from X-COP \citep{Eckert_2017}, rescaled to comoving densities. These profiles assume the functional form by \citet{Vikhlinin_2006} but drop the additional core term and thus present a monotonically steepening slope, in contrast to our profiles at $z \approx 2$ that also steepen towards the centre. The ubiquity of double-$\beta$-shaped profiles at high-$z$ is not surprising, as cooling times scale as $\tau_\mathrm{cool} \propto 1/\rho$ and dynamical timescales (hence, merger rates) as $\tau_\mathrm{dyn} \propto 1/\sqrt{\rho}$, making cooling more effective at high redshift, where physical baryonic densities are higher. Despite the different shapes, our central comoving densities are consistent with those of local CC clusters.\footnote{This is to be expected since central entropies, $K_e = k_B T n_e^{-2/3}$, are generally low. At $r=0.01 R_\mathrm{500c}$, $50\%$ ($90\%$) of our systems have ${K_e < 10 \, \mathrm{keV \, cm^2}}$ ($30 \, \mathrm{keV \, cm^2}$).}

At intermediate radii, the comoving electron densities of \textsc{Magneticum} protoclusters fall a factor $1.5-2$ below those of observed local clusters. This is in part contributed by the low ionisation fractions in $\lesssim(0.1-0.5)R_\mathrm{500c}$, but this effect alone ($\lesssim 20\%$ low with respect to full ionisation at $r \sim 0.5R_\mathrm{500c}$) cannot explain the differences in the $n_e(r)$ profiles and thus points at an actual redistribution of gas within the halo (e.g. by AGNs at lower redshift) as well as lower gas fractions in protoclusters when compared to low-redshift systems (see Fig.~\ref{fig:gas_depletion_fraction} for our simulated protoclusters, and figures 1 and 2 of \citealp{Eckert_2019} for the X-COP sample).

\subsection{Dependences on assembly state and central activity}
\label{s:results.other_dependencies}

\begin{figure*}
    \centering
    \begin{minipage}[c]{0.33\textwidth}
        \centering
        \includegraphics[width=\linewidth]{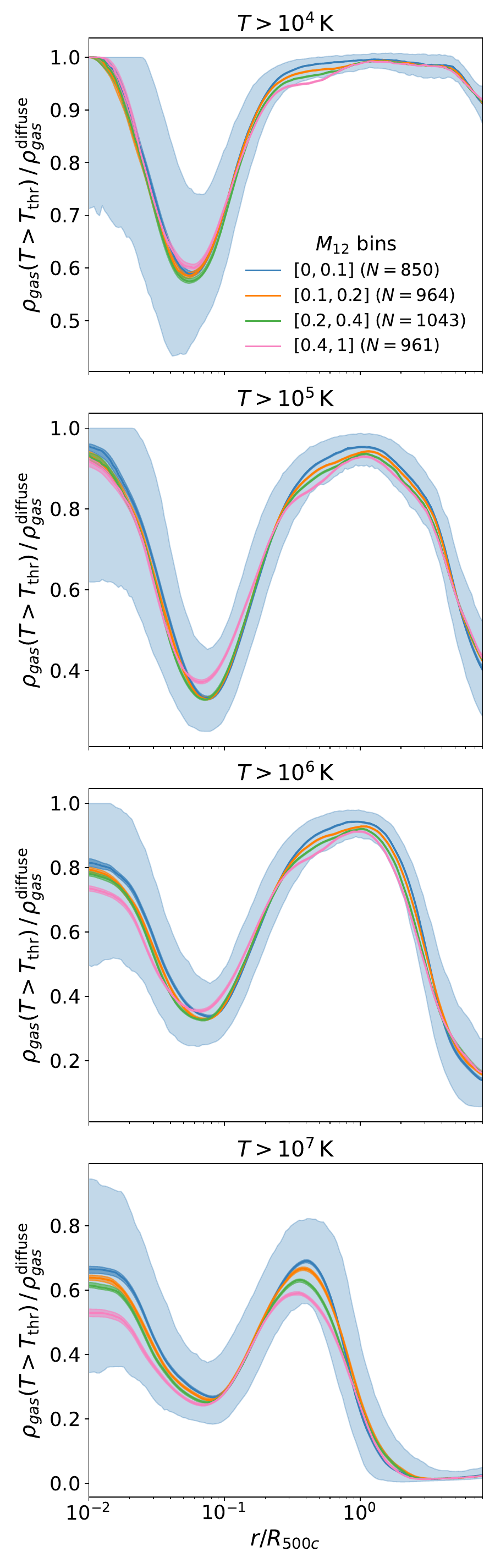}
    \end{minipage}
    \hfill
    \begin{minipage}[c]{0.33\textwidth}
        \centering
        \includegraphics[width=\linewidth]{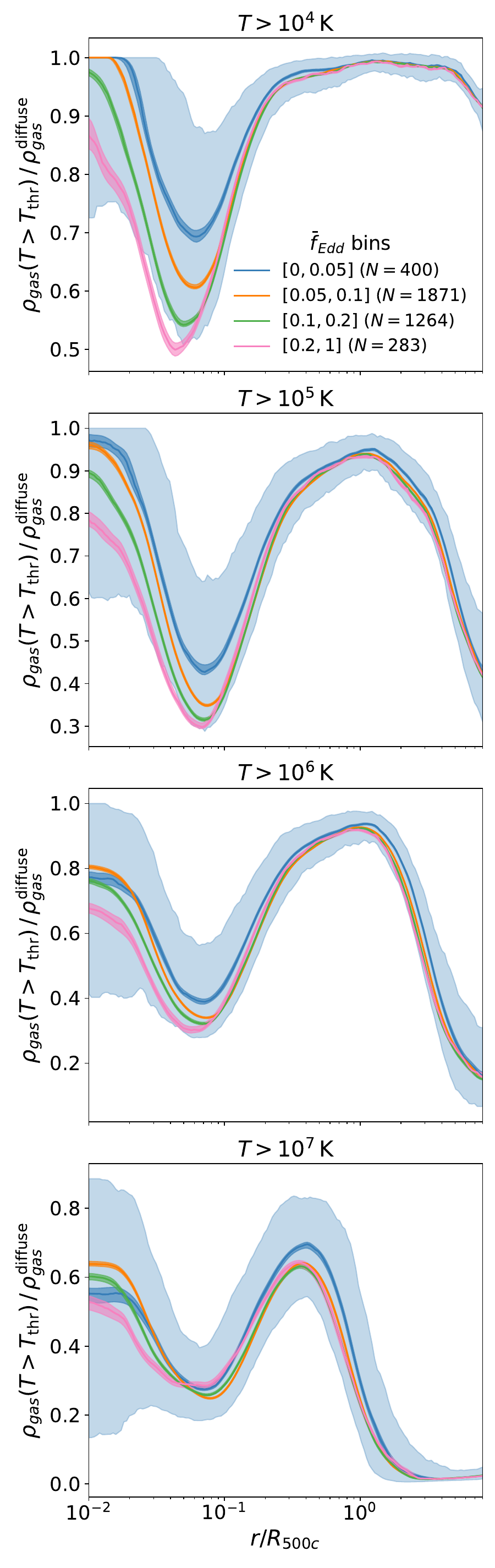}
    \end{minipage}
    \hfill
    \vrule width 0.4pt
    \hfill
    \begin{minipage}[c]{0.3\textwidth}
        \raggedbottom
        \centering
        \includegraphics[width=\linewidth]{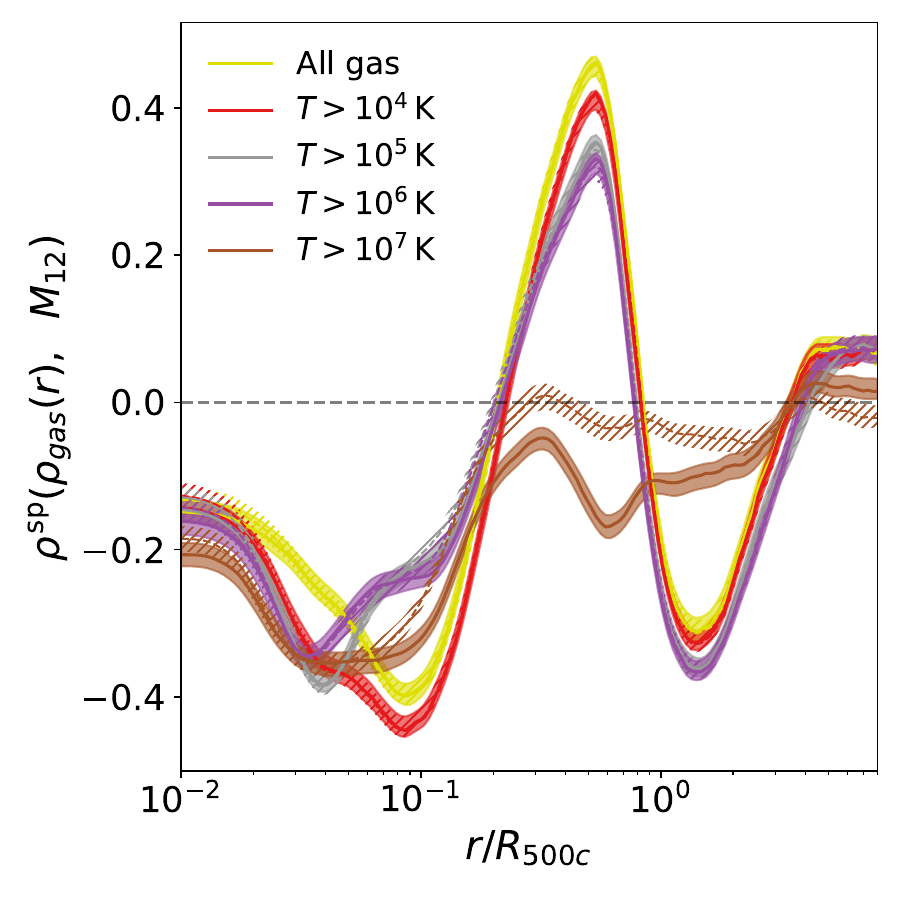}
        \includegraphics[width=\linewidth]{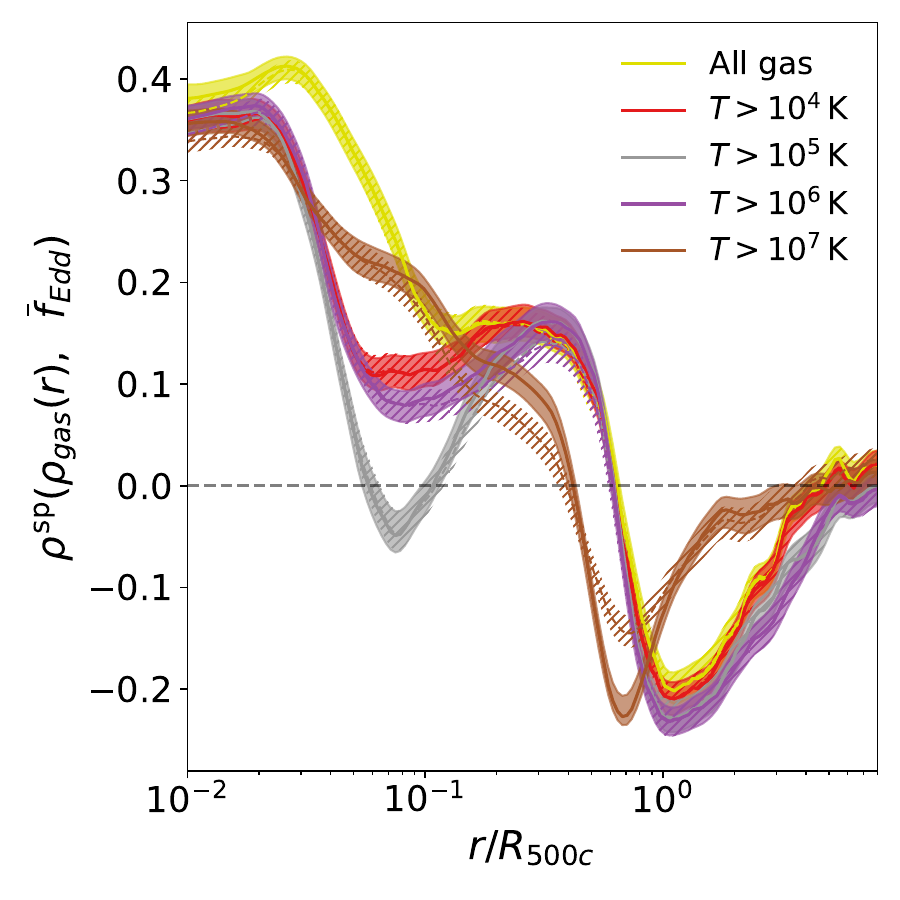}
    \end{minipage}
    \caption{Density fraction of diffuse gas above several temperature thresholds stacked by stellar mass ratio $M_{12}$ (indicator of assembly state, left column) and by average Eddington ratio $\bar f_\mathrm{Edd}$ (indicator of central activity, middle column). The four vertical panels in each column are equivalent to the second to fifth panels of Fig.~\ref{fig:gas_density_fractions_T}, normalised by the first one.
    The right panels show the Spearman rank-correlation coefficients between $M_{12}$ (top) and $\bar f_\mathrm{Edd}$ (bottom) and the density profiles, corrected for the mass dependence (solid; e.g. \citealp{Valles-Perez_2026}) or not (hatched).}
    \label{fig:gasfractions_and_correlationsassembly}
\end{figure*}

Beyond the overall depth of the potential well, set by $M_\mathrm{500c}$, the thermodynamic state of the ionised gas is also shaped by the protoclusters assembly history and central activity. Here we examine these dependences on the mass ratio $M_{12}$ and Eddington ratio $\bar f_\mathrm{Edd}$, both introduced in Sect.~\ref{s:methods.other_properties}. The analysis is summarised in Fig.~\ref{fig:gasfractions_and_correlationsassembly}, by studying the density fractions above several temperature thresholds in profiles stacked by $M_{12}$ (left column) and $\bar f_\mathrm{Edd}$ (middle column) and the correlations of these quantities with gas density above the same thresholds (right column). To better highlight the trends, we focused on the fractions of the diffuse gas density (see Eq.~\ref{eq:hotgas} instead of on the fractions of the total gas density shown in Fig.~\ref{fig:gas_density_fractions_T}).

The main effect of higher $M_{12}$, indicating a recent merger, on the warm ($T>10^4 \, \mathrm{K}$) and ionised ($T>10^5 \, \mathrm{K}$) gas density profiles amounts to lowering their mass fraction, with respect to the total diffuse gas, at intermediate radii ($[0.2, \, 1] R_\mathrm{500c}$), as the newly accreted gas is not yet fully thermalised within the protocluster potential well (as similarly pointed out in low-redshift clusters by simulations; \citealp{Lau_2015, Chen_2019, Valles-Perez_2026}). At the temperatures of typically X-ray emitting plasma ($10^6$, $10^7 \, \mathrm{K}$), there is the additional effect of slightly decreasing the hot gas fractions in the core regions, as major mergers most often disturb the core and promote mixing \citep{ZuHone_2011}, and can also decrease central densities \citep{Valles-Perez_2026}, halting accretion onto the central AGNs \citep[e.g.][]{Chadayammuri_2021}. 

The correlations between gas densities above each of these thresholds and $M_{12}$ are shown in the top-right panel of Fig.~\ref{fig:gasfractions_and_correlationsassembly}. When combined with the total gas density profile (Sect.~\ref{s:results.density_mass}), these results yield a moderate positive correlation ($\rho^\mathrm{sp} \sim 0.4$) between $M_{12}$ and densities above all temperature thresholds (except $10^7 \, \mathrm{K}$) at intermediate radii. The strength of the correlation gradually decreases with increasing $T^\mathrm{thr}$. For $\rho_\mathrm{gas}(T>10^7 \, \mathrm{K})$, the decrease in the hot-gas fraction with $M_{12}$ compensates the increase of the total density profile. As a result, the density profile of the bremsstrahlung-emitting proto-ICM is fairly independent of $M_{12}$. At central radii, the correlations are overall weaker ($\rho^\mathrm{sp} \sim - 0.2$) and show little dependence on $T^\mathrm{thr}$.

The integrated Eddington ratio, $\bar f_\mathrm{Edd}$, which quantifies the relative strength of AGN feedback, is expected to bear its strongest effect in the protocluster centre. Indeed, the bottom-right panel of Fig.~\ref{fig:gasfractions_and_correlationsassembly} shows that the correlations of the gas profiles with $\bar f_\mathrm{Edd}$ peak at central radii, with values of $\sim 0.4$. The gas fractions above different temperature thresholds (middle column) reveal a clear association of $\bar f_\mathrm{Edd}$ with the depth of the dip at $\sim 70 \, \mathrm{kpc}$ discussed in the previous sections, particularly at the lowest temperature thresholds. Protoclusters with a strongly accreting central AGN (mainly in the quasar mode) host a very large ($\sim 50\%$) fraction of neutral gas outside the region immediately heated by thermal AGN feedback. This effect progressively weakens for bins of smaller $\bar f_\mathrm{Edd}$, supporting the association of this feature to AGN feedback.

We checked that this feature occurs at radii substantially larger than the characteristic numerical scales of the simulation. In particular, its location ($r \sim 70 \, \mathrm{ckpc}$) lies over an order of magnitude beyond the gravitational softening length, and typically several central SPH smoothing lengths (which also mark the scale where AGN energy is coupled to the gas). At the same time, the feature lies roughly within the transition region between the central ISM (typically dominant within $r \lesssim 40 \, \mathrm{kpc}$, and staying with mass fractions over $10\%$ up to $r \sim 100 \, \mathrm{ckpc}$) and the hot proto-ICM. While these arguments disfavour a straightforward numerical origin, a dedicated convergence study with varying resolution and comparison with alternative AGN feedback models (both thermal and kinetic) would be required to establish the physical nature of this feature more firmly.

It is also worth discussing the sign of the correlations in the centre. Higher $\bar f_\mathrm{Edd}$ correlates positively with central gas densities, despite AGN feedback acting to heat and remove gas from the centre. While high instantaneous SMBH accretion rates require high central densities, the accretion rates considered here are averaged over a large ($\sim 500 \, \mathrm{Myr}$) timescale. The persistence of a positive correlation between $\bar f_\mathrm{Edd}$ and central densities implies that, within the \textsc{Magneticum} feedback model, SMBHs experience coherent accretion modes over extended timescales, rather than rapidly alternating between higher and lower accretion modes.

\begin{figure*}
    \centering
    \includegraphics[width=0.33\linewidth]{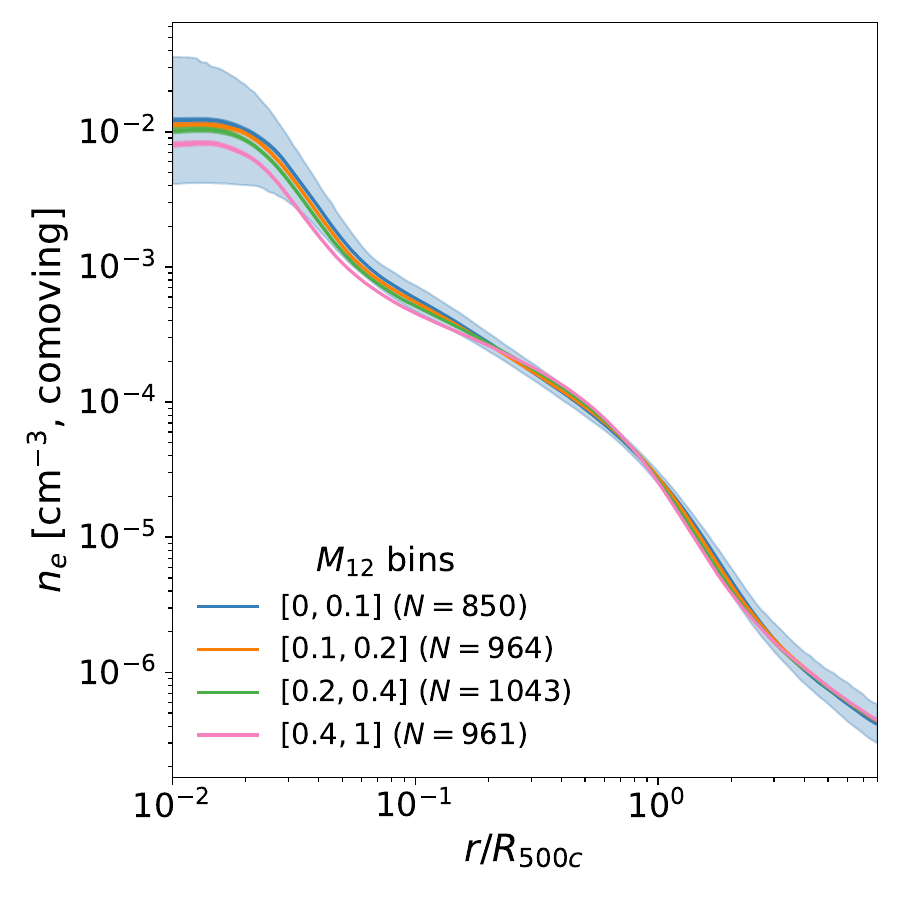}~
    \includegraphics[width=0.33\linewidth]{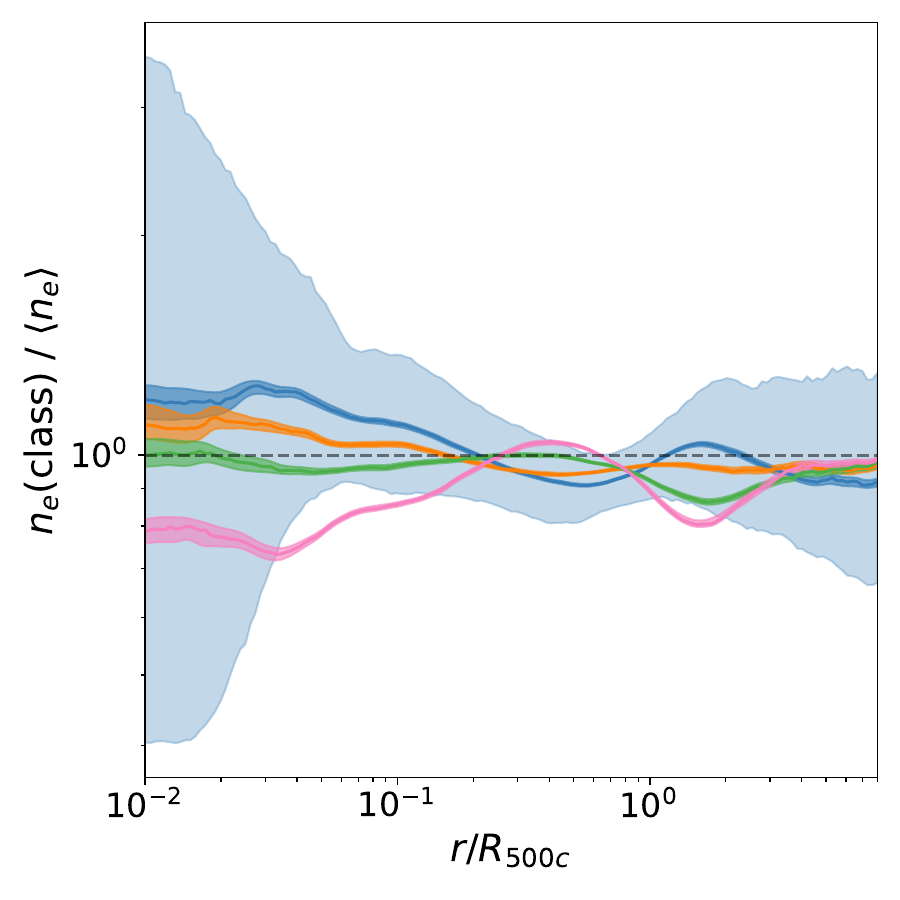}~
    \includegraphics[width=0.33\linewidth]{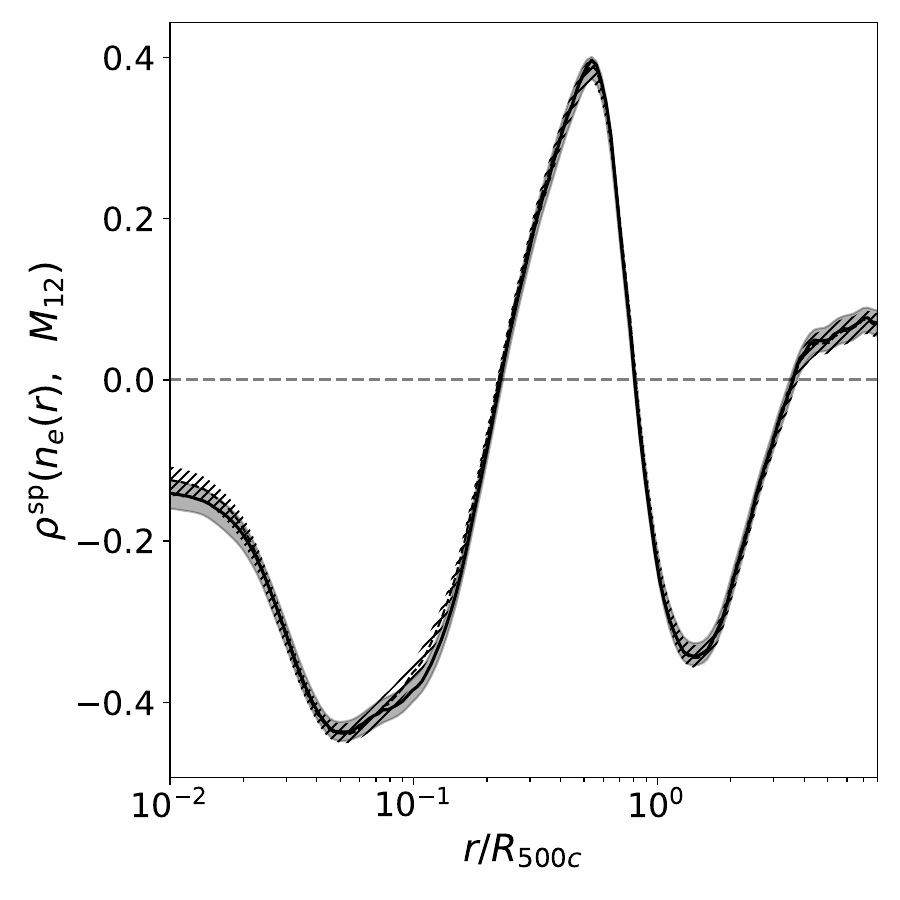}
    \includegraphics[width=0.33\linewidth]{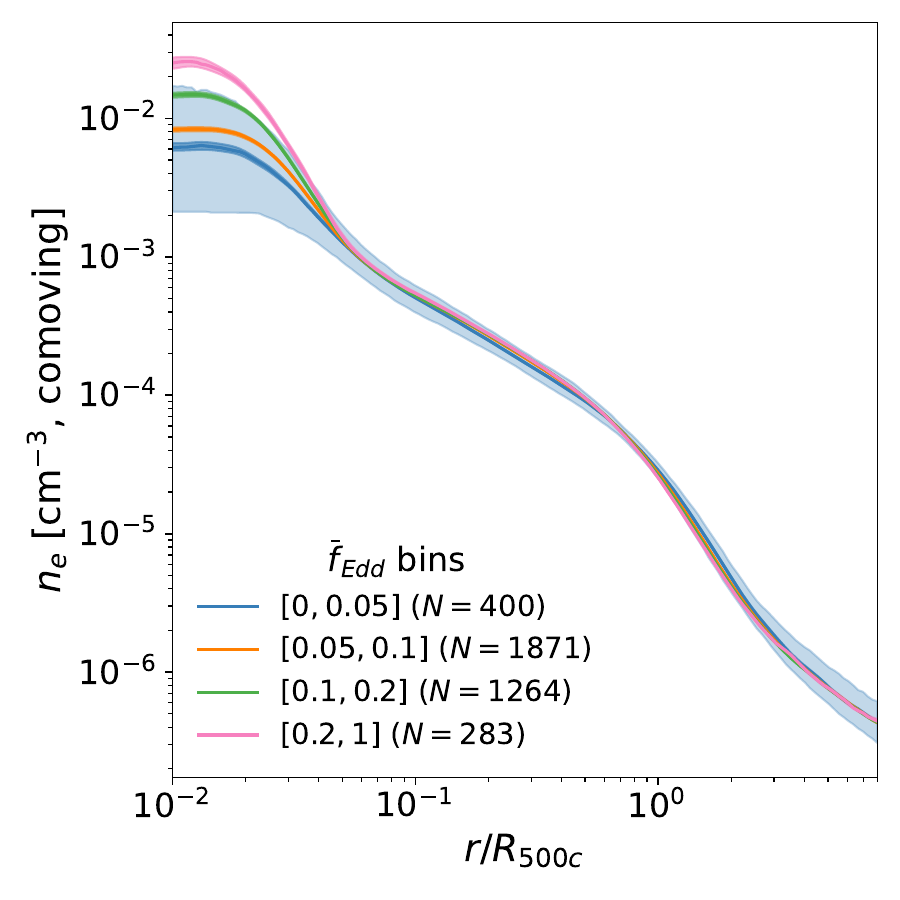}~
    \includegraphics[width=0.33\linewidth]{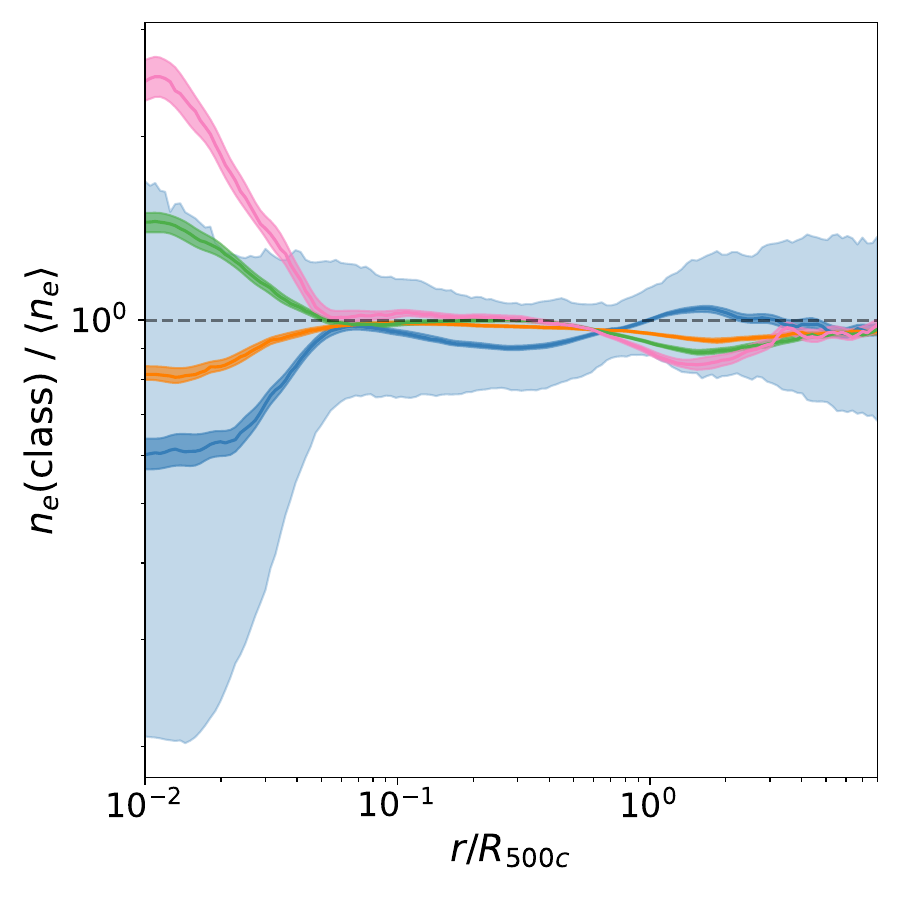}~
    \includegraphics[width=0.33\linewidth]{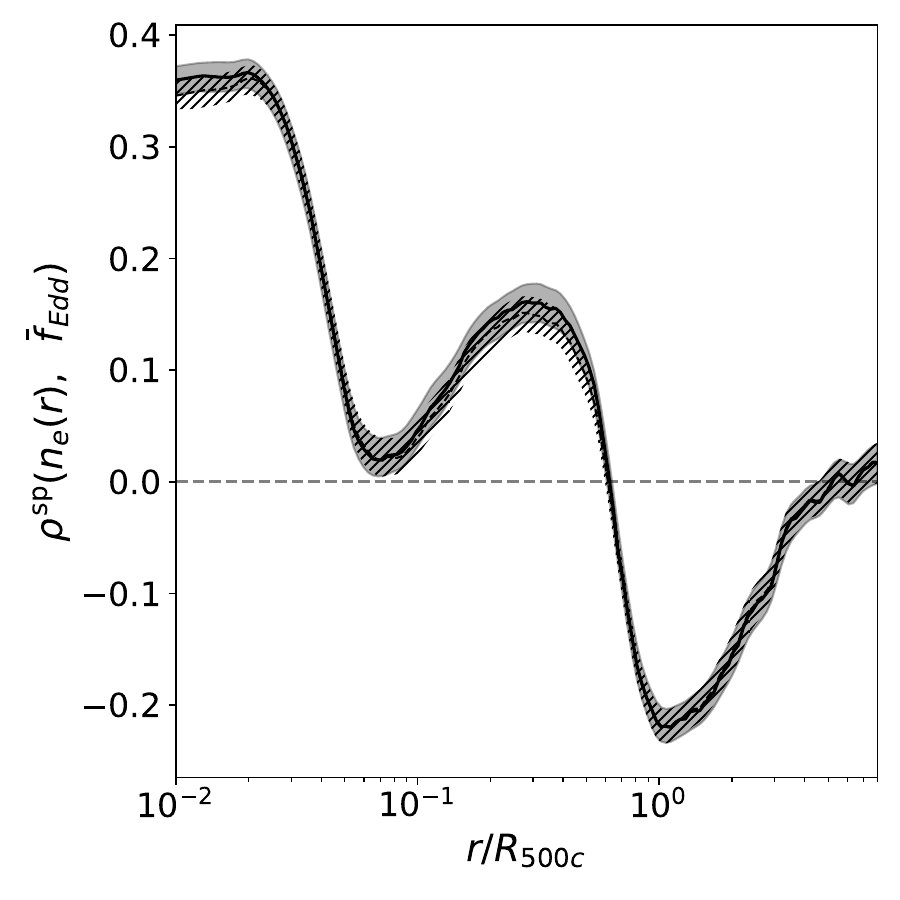}
    \caption{Analysis of the dependence of the electron number density profiles, $n_e(r)$, on assembly state (stellar mass ratio $M_{12}$, upper row) and central activity (average Eddington ratio, $\bar f_\mathrm{Edd}$, lower row). \textit{Left:} Electron comoving number density profiles. \textit{Middle:} Same as left panel but all profiles have been normalised by the stack over the whole sample. \textit{Right:} Spearman rank correlation between the $n_e(r)$ profiles at each $r$ and the relevant variable.}
    \label{fig:electrondensity_othervariables}
\end{figure*}

Finally, in Fig.~\ref{fig:electrondensity_othervariables}, we summarise the effect of $M_{12}$ (top row) and $\bar f_\mathrm{Edd}$ (bottom row) on the $n_e(r)$ profiles. The left, middle, and right columns are equivalent to the top-left, top-right and bottom-right panels of Fig.~\ref{fig:electrondensity_mass}, respectively. Altogether, the effect of increasing $M_{12}$ (i.e. moving from relaxed to disturbed protoclusters; top row) corresponds to a displacement of ionised gas from the centre to intermediate radii. Even though disturbed protoclusters have lower fractions of hot gas (see Fig.~\ref{fig:gasfractions_and_correlationsassembly}, left column), this effect is subdominant compared to the effect on the total gas density. The magnitude of the variation ranges from a $\sim 50\%$ relative difference (between the lowest and highest bins in $M_{12}$) in the centre, to a $\pm 10\%$ effect at $r \in [0.2, \, 0.8] R_\mathrm{500c}$, where the scatter is much lower. Correlations for both effects reach $|\rho^\mathrm{sp}| \sim 0.4$.

As for the dependences on AGN feedback strength ($\bar f_\mathrm{Edd}$; lower panels of Fig.~\ref{fig:electrondensity_othervariables}), as expected, the most remarkable effect is found in the cores ($r \lesssim 0.05 R_\mathrm{500c}$), where the protocluster samples built from the lowest- and highest-accreting SMBHs depart by a factor of $\gtrsim 4$, with a positive correlation of $\bar f_\mathrm{Edd}$ with central densities reaching $\rho^\mathrm{sp} \sim 0.4$. In this case, the correlations observed at larger radii (especially the ones at $\sim R_\mathrm{500c}$) are likely to be mostly contributed by compensation (increased central gas densities implying lower ones at higher $r$), and the influence of the gas distribution on the determination of $R_\mathrm{500c}$.

\section{Discussion}
\label{s:discussion}

This section further discusses the previous results on the protocluster thermal structure. This includes the validity of our assumptions to derive electron densities (Sect.~\ref{s:discussion.assumptions}), simple estimates on their X-ray emission (Sect.~\ref{s:discussion.xray}), comparisons with the Spiderweb protocluster (Sect.~\ref{s:discussion.spiderweb}) and the scatter around the $n_e(r)$ profiles (Sect.~\ref{s:discussion.all_together}).

\subsection{Validity of the assumptions to derive $n_e$}
\label{s:discussion.assumptions}

In Sect.~\ref{s:methods.postprocessing}, we described our approach to extract electron densities from the simulation data relying on two key assumptions. Below, we briefly justify below their validity.

The first assumption corresponds to neglecting the effect of photoionisation and therefore exclusively accounting for the effects of collisional ionisation. While this choice is motivated by simplicity, here we discuss why its effects on our results are fairly limited. First, while photoionisation is not explicitly accounted for in our post-processing computation, the thermodynamic state of the gas is affected by the associated photoheating, which is included in the simulation. Specifically regarding photoionisation by the metagalactic UV background, the high central physical densities imply that gas in the inner regions ($r \lesssim [0.1-0.2]R_\mathrm{500c}$) is expected to be largely self-shielded \citep[see e.g.][]{Rahmati_2013}. Therefore, this effect is unlikely to affect our main results in this region. At larger radii, instead, where self-shielding is ineffective, the gas is predominantly collisionally ionised, and the impact of photoionisation remains subdominant. 

The role of a local radiation field from the AGN is far more uncertain. The low ionisation fractions reported at $r \sim 70 \, \mathrm{kpc}$ (see Fig.~\ref{fig:electrondensity_mass}) lie in a region where photoionisation from the AGN could have some effect, locally increasing the ionisation fraction. However, it is not feasible at this stage to estimate in how far it would attenuate this feature, since the possible effect depends strongly on the detailed properties of the AGN (its luminosity, spectrum, duty cycle, isotropy), which we do not model here. 

Finally, to assess the possible effect of non-equilibrium ionisation, we estimated the CIE timescales, $\tau_\mathrm{ion}$, using the rate coefficients from \citet{Voronov_1997}. In general terms, we find them to be four to six orders of magnitude below the dynamical and cooling timescales in the central regions. This suggests that CIE is an excellent approximation for the bulk of baryonic mass in protocluster cores. While this does not rule out that departures from equilibrium may still occur locally in shocks or very rapidly cooling clumps, it justifies our usage of this assumption for estimating the radially-averaged profiles of $n_e(r)$. Overall, these considerations support the validity of our assumptions for deriving radially averaged $n_e(r)$ profiles in protocluster core environments.

\subsection{Estimates on X-ray emission from protoclusters}
\label{s:discussion.xray}

Despite the fact that some determinations of proto-ICM densities have been achieved in the recent years \citep{Tozzi_2022, Lepore_2024}, the faintness of their signal generally makes the task prohibitive for current X-ray telescopes. To put our results into context and motivate the need for calibrating the density profiles of protoclusters with simulations, we present here a simplified analysis. This is aimed, not at reproducing the actual X-ray emission from protoclusters, but just to obtain a rough estimate on the scaling of the exposure times required to detect their emission from several radial annuli.

\begin{figure}
    \centering
    \includegraphics[width=\linewidth]{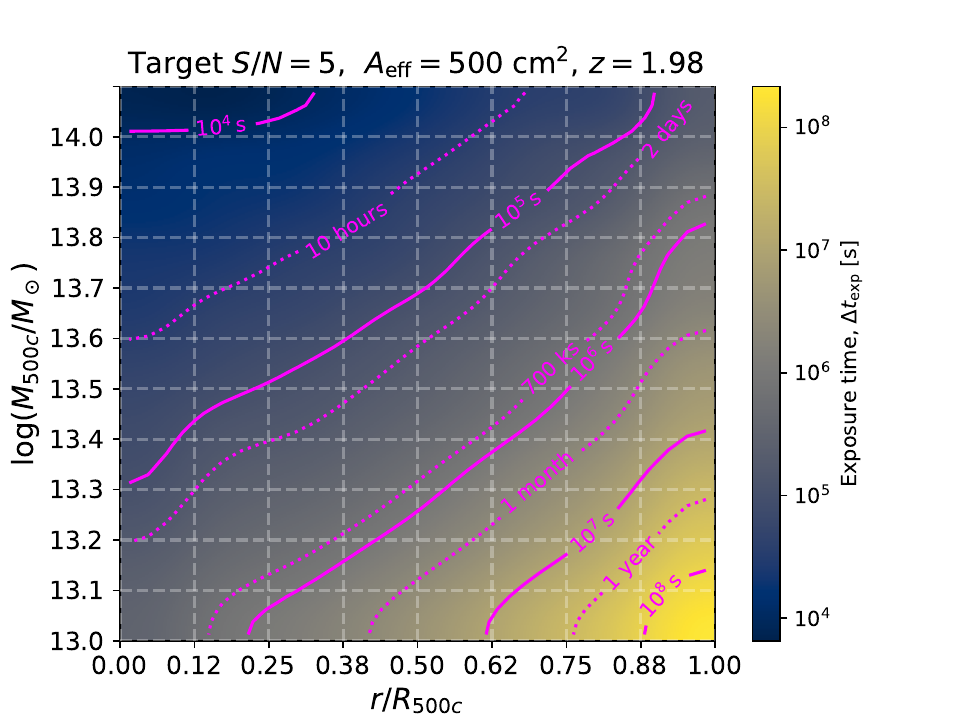}
    \caption{Estimation of the exposure time, $\Delta t_\mathrm{exp}$, to reach a signal-to-noise ratio of $\mathrm{S/N}=5$, with an effective area of $A_\mathrm{eff} = 500 \, \mathrm{cm}^2$, from the typical proto-ICM at $z\approx 2$. The colour map shows our predicted $\Delta t_\mathrm{exp}$ for different protocluster masses (vertical axes) and within eight radial bins (equally spaced from $0$ to $R_\mathrm{500c}$; horizontal axis). The colour map has been smoothed for visualisation purposes. Solid and dashed magenta lines indicate some representative values of $\Delta t_\mathrm{exp}$ for better reference.}
    \label{fig:xray}
\end{figure}

The details for our modelling of the X-ray emission from protoclusters are described in App. \ref{s:app.xray_method}, together with some examples on the resulting X-ray surface brightness maps. Using them, we obtained a simplistic, order-of-magnitude estimate of the exposure time that would be required in order to detect the diffuse X-ray emission from the proto-ICM in the $[0.5, \, 2] \, \mathrm{keV}$ band in several radial annuli (all with $\Delta r = R_\mathrm{500c} / 8$). We target a signal-to-noise ratio of $\mathrm{S/N} = 5$ as a reasonable threshold for the sole detection of this emission, assuming an average background rate of $b = 5 \times 10^{-4} \, \mathrm{cts \, s^{-1} \, {arcmin}^{-2}}$ and an effective area of $A_\mathrm{eff} = 500 \, \mathrm{cm}^2$. The results of such an estimation are summarised in Fig.~\ref{fig:xray}, as a function of $M_\mathrm{500c}$ and radial bins.

This analysis highlights that, for X-ray telescopes similar to \textsc{Chandra} and with optimistic assumptions (e.g, given its progressive loss of sensitivity in the soft band; \citealp{Plucinsky_2018}), reasonable exposure times ($\sim 100 \, \mathrm{ks}$) would only be able to detect emission at large radii from the most massive (and rarest) protoclusters, while in order to target lower-mass, more abundant protoclusters, exposure times quickly rise to months or years. 
In the figure, we also include, for reference, a contour line corresponding to the $700 \, \mathrm{ks}$ exposure that allowed the measurement of an average electron number density in the Spiderweb protocluster \citep[][with $M_\mathrm{500c} \approx 3 \times 10^{13} M_\odot$ at $z_\mathrm{Spdw} = 2.16$]{Tozzi_2022}.
They detected thermal X-ray emission from this system out to a radius of $100 \, \mathrm{kpc} \approx 0.44 R_\mathrm{500c}$. Consistently, our optimistic estimation predicts achieving a detection with $\mathrm{S/N}=5$ out to $\sim 0.6 R_\mathrm{500c}$, serving as a validation of our forecast.

The high effective area of future X-ray facilities, such as \textsc{NewAthena} \citep{Cruise_2025}, will offer the possibility to significantly improve the detection of faint high-$z$ emission. However, their comparatively coarser angular resolution ($\sim 10 \, \mathrm{arcsec}$) will make it exceedingly challenging to disentangle diffuse proto-ICM emission from compact sources. Furthermore, while we predict exposure times of the order of $10^{4-8} \, \mathrm{s}$ to achieve significant detections of different regions of the proto-ICM, retrieving densities through spectral fitting requires higher-quality data and hence considerably higher exposure times capable of providing enough net counts in each radial bin.

It is thus unlikely that X-ray observations in the next decades will be able to place strong constraints on proto-ICM densities. While SZ observations also provide a promising channel to study the thermal state of this plasma, deriving densities would require assumptions on its temperature and ionisation structure. In this context, cosmological simulations, which have shown considerable success in reproducing properties of lower-$z$ galaxies and clusters, are essential tools to provide predictions on the proto-ICM structure for a broad range of applications, including the study of its magnetism or synergies with SZ observations.

\subsection{Comparison with Spiderweb measurements}
\label{s:discussion.spiderweb}

\begin{figure}
    \centering
    \includegraphics[width=0.9\linewidth]{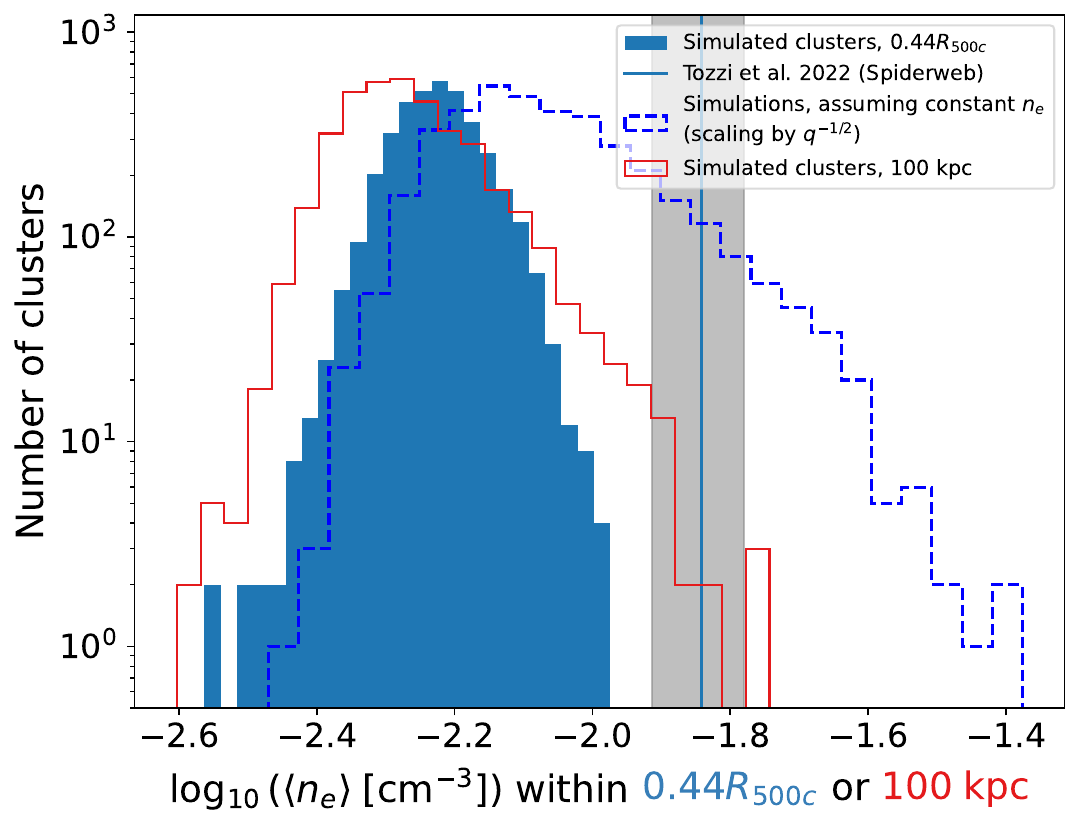}
    \includegraphics[width=\linewidth]{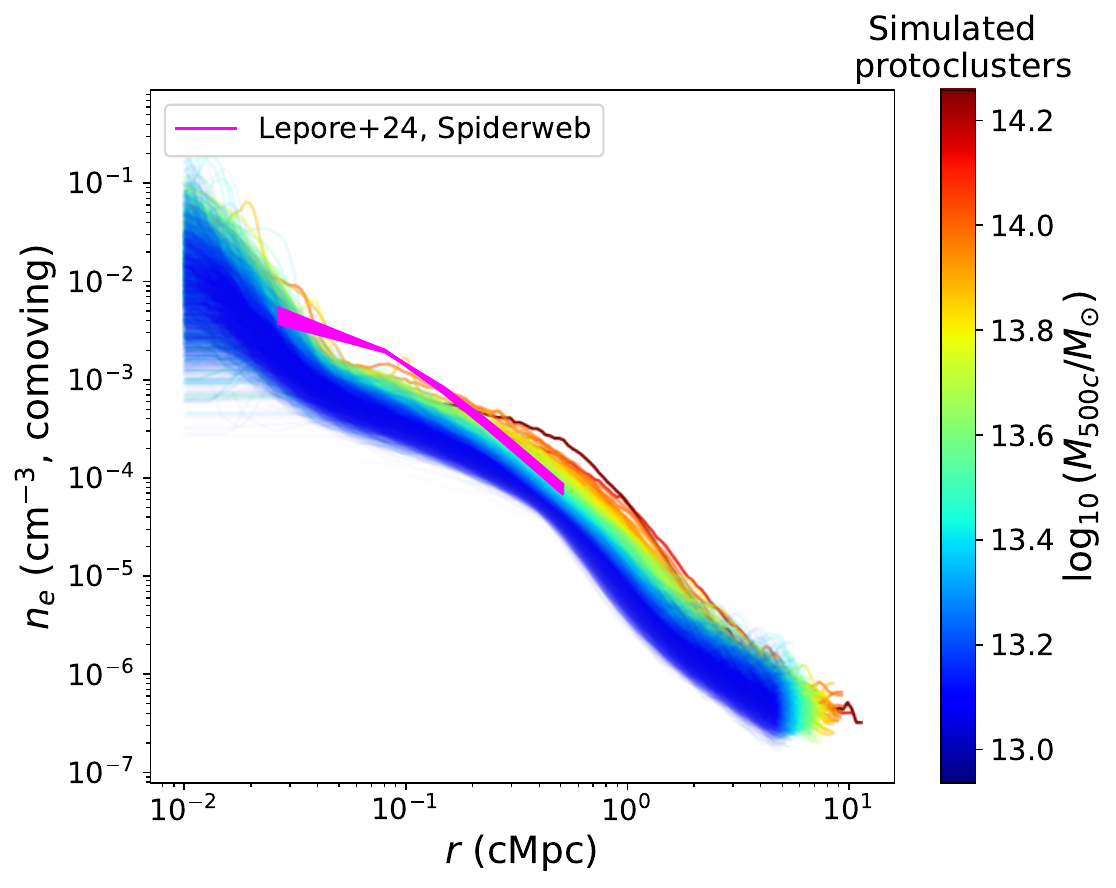}
    \caption{Comparison of simulated $n_e(r)$ to data observed for the Spiderweb protocluster. \textit{Top panel:} Simulated $\langle n_e \rangle$ within $0.44 R_\mathrm{500c}$ (blue filled histogram) together with the corrected value accounting for the observational assumption of homogeneous distribution ($\propto q^{-1/2}$, blue dashed lines) in comparison to the value measured by \citet{Tozzi_2022}. The red histogram instead shows  the mean densities within a physical aperture of $100 \, \mathrm{kpc}$. \textit{Bottom panel}: Comoving $n_e(r)$ profiles, colour-coded by $M_\mathrm{500c}$, and the profile derived for \citet{Lepore_2024} for the Spiderweb system (magenta). }
    \label{fig:compare_2_spiderweb}
\end{figure}

From their $700 \, \mathrm{ks}$ Chandra exposure of the Spiderweb central region, \citet{Tozzi_2022} report a mean density ${\big\langle n_e \big\rangle_\mathrm{Spdw} = (1.51 \pm 0.24  \pm 0.14) \times 10^{-2} \, \mathrm{cm}^{-3}}$ within an aperture of $100 \, \mathrm{kpc} \approx 0.44 R_\mathrm{500c}^\mathrm{Spdw}$. In the top panel of Fig.~\ref{fig:compare_2_spiderweb}, we compare this value (vertical line with grey shaded region for its uncertainty) with the electron densities, averaged inside $0.44 R_\mathrm{500c}$, of our simulated protoclusters (blue, filled histogram), which follow a relatively narrow, nearly log-normal distribution with ${\langle \log_{10} \left( n_e/\mathrm{cm^3} \right) \rangle = -2.22 \pm 0.07}$. This suggests a moderate (${\sim 3 \sigma}$) tension with the value observed for the Spiderweb protocluster in this fiducial comparison, and as a matter of fact, none of our $3818$ analysed systems reaches such high averaged densities in a comparable aperture. This result is consistent with the fact that the spatially-resolved profiles of \citet{Lepore_2024} lie systematically above those of our simulated systems (see grey triangles in the upper left panel of Fig.~\ref{fig:electrondensity_mass}) ranging from a factor of $\sim 3$ (in the centre) to $+20\%$ (towards $\sim 0.5 R_\mathrm{500c}$).

The possibility that some of the current generation of cosmological simulations may underpredict the central densities of some high-$z$ systems has already been suggested in the recent literature (see e.g. \citealp{Travascio_2025}). In particular, the purely thermal nature of the AGN feedback implementation in \textsc{Magneticum} may represent, perhaps, the most important limitation affecting this comparison. Different works have discussed how thermal energy injection alone --at similar efficiencies as implemented in kinetic feedback models-- is unable to offset cool cores \citep{Barai_2016}, while stronger thermal feedback tends to become overly ejective \citep{Gaspari_2014}. Therefore, the discrepancy found above could be qualitatively consistent with a known limitation of the subgrid modelling in these simulations.

To assess the robustness of this discrepancy to our analysis, we explored two sources of systematic uncertainties affecting the comparison. First, the average density measurement of \citet{Tozzi_2022} relies on the assumption of a constant density, which may bias the inferred densities high. To estimate the impact of this assumption, we compared the X-ray luminosity of our inhomogeneous ICMs to equivalent versions containing the same gas mass and temperature distribution, but with a uniform $n_e$. We then propagated the luminosity ratio, $q \equiv  L_\mathrm{X,homog} / L_\mathrm{X} \lesssim 1$, into a density correction, $\langle n_e \rangle^\mathrm{corr} = \langle n_e \rangle q^{-1/2}$, based on a simple scaling argument, $L_X \propto n_e^2$. Under this correction, a non-negligible fraction of systems reaches mean corrected $n_e$ comparable to or higher than the values measured for the Spiderweb in the same aperture ($\gtrsim 10\%$; see the dashed blue histogram in the upper panel of Fig.~\ref{fig:compare_2_spiderweb}). While accounting for this effect would point in the direction of easing the tension, it is also present when we compare with the profiles derived by \citet{Lepore_2024}, which drop the assumption of constant $n_e$. This suggests that this effect alone does not justify, in the case of the Spiderweb, the discrepancy.

Second, the fiducial comparison is anchored to the value of $R_\mathrm{500c}$, which itself depends on the value of $M_\mathrm{500c}$, which has been somewhat debated in the literature \citep[see][]{Kurk_2004, Miley_2006}. Repeating the comparison with a fixed physical aperture of $100 \, \mathrm{kpc}$ instead of $0.44 R_\mathrm{500c}$ (red histogram) would likewise increase the level of agreement. A complementary view of this is shown in the lower panel of Fig.~\ref{fig:compare_2_spiderweb}, where we show the simulated profiles (colour-coded by $M_\mathrm{500c}$) and the one derived by \citet[][shown as a pink band]{Lepore_2024} in physical coordinates, rather than scaled by $R_{500c}$. The discrepancy is softened when the observed profile is compared to our most massive systems (redder lines; see also App.~\ref{s:app.spiderweb} for a quantitative analysis). These objects, however, exceed the Spiderweb's $M_\mathrm{500c}$ value derived from X-ray and SZ scaling relations \citep{Tozzi_2022, DiMascolo_2023} by a factor of $\sim 3$; so that, while the agreement of the electron densities improves in this regime, it is unlikely that this can alone resolve the discrepancy.

Overall, the fiducial comparison indicates that the \textsc{Magneticum} simulations do not reproduce the high central densities observed in the Spiderweb halo. At the same time, the systematic effects discussed above substantially affect the quantitative level of agreement, making it difficult to assess the statistical significance of such a discrepancy within the scope of this work. Finally, it is worth noting that the Spiderweb may itself be an unusual object at $z \simeq 2$, mainly studied due to its strong radio beacon in clear interaction with the surrounding medium. The rarity of such an object in our simulations may partly reflect this selection effect (further backed by the correlation between central densities and nuclear activity discussed in Sect.\ref{s:results.other_dependencies}). 

\subsection{Combined constraining power of multiple parameters}
\label{s:discussion.all_together}

To quantitatively estimate in how far protocluster mass ($M_\mathrm{500c}$), as well as $M_{12}$ and $\bar f_\mathrm{Edd}$, reduce the scatter in the electron density profiles, we employ a variance-decomposition approach, similar to the classical ANOVA \citep{Fisher_1970}. In essence, we discretise each of the three control variables in $n_b = 8$ bins with equal counts, for a total of $n_b^3 = 512$ groups, and recursively merge neighbouring bins which do not contain at least five objects to ensure accurate variance calculations. We checked that the results we extract below are largely insensitive to moderate variations in $n_b \in [3 , 10]$.

For each group $g$ (chosen by fixing one, two, or three variables), we stacked the individual $(n_e)_i(r)$ profiles to obtain $\langle n_e \rangle_g(r)$. From this, we computed the between-groups sum of squares:
\begin{equation}
    \mathrm{BSS}(r) = \sum_g N_g \left(\langle \log_{10} n_e (r) \rangle_g - \langle \log_{10} n_e (r)\rangle_\mathrm{all}  \right)^2,
\end{equation}
\noindent with $N_g$ as the number of protoclusters in the group $g$ and $\langle \log_{10} n_e (r) \rangle_\mathrm{all}$ as the profile stacked over the whole sample. Then, we defined the fraction of variance explained as
\begin{equation}
    \eta^2(r) = \frac{\mathrm{BSS}(r)}{\sum_{i} \left( \log_{10} \left[(n_e)_i(r)\right] - \langle \log_{10} n_e (r)\rangle_\mathrm{all}\right)^2},
\end{equation}
\noindent where the denominator is the total sum of squares. In this way, $\eta^2(r)$ is parallel to the $\eta^2$ statistic used in the classical ANOVA formulation without any parametric assumption.

\begin{figure}
    \centering
    \includegraphics[width=0.9\linewidth]{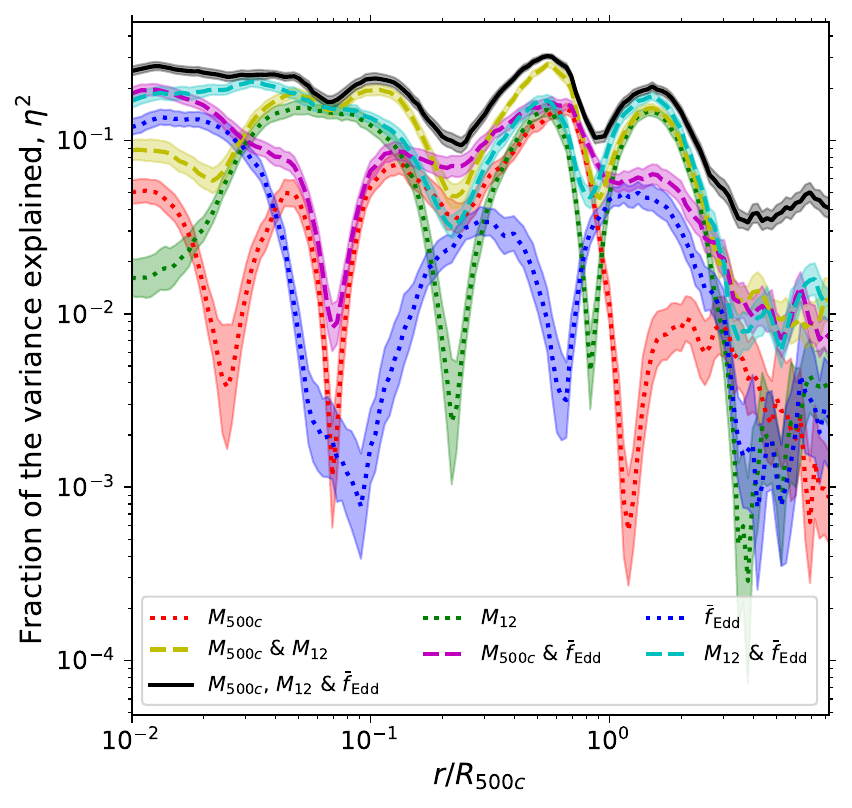}
    \caption{Results from the variance decomposition study for $n_e(r)$ as a function of protocluster mass $M_\mathrm{500c}$, mass ratio $M_{12}$, and averaged Eddington ratio $\bar f_\mathrm{Edd}$. Each line represents the radial profile of the fraction in the variance of $n_e(r)$ that can be explained by each single variable (dotted lines), pair of variables (dashed lines), or all three variables (solid line).}
    \label{fig:anova}
\end{figure}

These results are shown, for individual variables (dotted lines), pairs of variables (dashed lines) and all three variables together (solid line) in Fig.~\ref{fig:anova}. Focussing on single variables, we find that $\bar f_\mathrm{Edd}$, measuring the impact of the central AGN, is the dominant contribution within $0.03 R_\mathrm{500c}$, while at higher radii, $M_{12}$ tends to account for a larger portion of the variability in $n_e(r)$. Mass reaches a similar predictive power on $n_e(r)$, especially at these intermediate radii ($0.1 \lesssim r /R_\mathrm{500c} \lesssim 1$).

Combining variables increases the amount of variance that can be explained. The AGN and assembly state ($\bar f_\mathrm{Edd}$ and $M_{12}$; at $r \lesssim 0.1 R_\mathrm{500c}$) or assembly state and mass (at $0.1 \lesssim r/R_\mathrm{500c} \lesssim 1$) reach $20\%$ of the variance in the profiles. The combination of the three reaches values of $\eta^2 \sim 30\%$. 

Therefore, the knowledge of mass, ratio of first satellite to BCG stellar masses, and the Eddington ratio of the central BH accounts for up to $\sim 30 \%$ of the logarithmic variance in the $n_e(r)$ profiles. The remaining $70\%$ of the variance cannot be statistically accounted for by these variables. There might be two fundamental contributions to this. On the one hand, and more importantly, the approach is based on studying spherically-averaged profiles from non-spherically symmetric objects. Therefore, a large portion of the object-to-object variance could be attributed to departures from sphericity, that introduce scatter to the values of $n_e$ at each given $r$. On the other hand, neither $M_{12}$ or $\bar f_\mathrm{Edd}$ are complete indicators of dynamical state or AGN activity, respectively. For instance, several works mainly focused at $z = 0$ \citep{Valles-Perez_2023, Haggar_2024, Valles-Perez_2025} showed how different indicators of assembly state bring complementary information about the cluster internal properties and evolution. Regarding $\bar f_\mathrm{Edd}$, our strategy for estimating it from a pair of snapshots imposes a timescale ($\sim 500 \, \mathrm{Myr}$). Despite the fact that we found coherent results by using either the integrated or the instantaneous Eddington ratios, the scatter is large, and it is generally not possible to capture the integrated effect of AGN feedback on the proto-ICM with a single quantity. Nevertheless, the complementarity of the different indicators investigated here motivate the calibration of electron density profiles as a function of these variables; a topic that will be pursued in a subsequent work.

\section{Conclusions}
\label{s:conclusions}

We have analysed the gas and electron number densities of a sample of over $3818$ protocluster cores -- i.e. the baryonic structures centred on (quasi-)virialised DM haloes -- at $z = 2$ with masses spanning from $10^{13} \, M_\odot$ to $2 \times 10^{14} \, M_\odot$ extracted from a $(640 \, h^{-1} \, \mathrm{Mpc})^3$ box from the \textsc{Magneticum} suite. Our main conclusions are as follows:

\begin{enumerate}
    \item The total gas density profiles deviate moderately from self-similarity at inner ($r \lesssim 0.1 R_\mathrm{500c}$) and intermediate ($r \sim [0.3-0.8] R_\mathrm{500c}$) radii. They are generally inconsistent with a single-$\beta$ model and instead steepen towards the centre, with the effect being stronger at higher masses. At intermediate radii, more massive clusters are on average more gas rich.
    \item The temperature structure of protoclusters is complex. Cold ISM dominates ($\sim 60\%$) at $r \lesssim 70 \, \mathrm{kpc}$ and does not scale self-similarly with mass. Temperatures consistent with full ionisation only reach significant fractions ($\sim 80\%$) at $(0.1-0.5) R_\mathrm{500c}$, generally occurring at larger radii for lower-mass systems. Only the most massive protoclusters reach temperatures where bremsstrahlung dominates the X-ray emission. 
    \item Electron number density profiles present a pronounced double-$\beta$ structure. The inner component steepens inside $r/R_\mathrm{500c} < 0.1$, and the effect becomes stronger with increasing mass. Ionisation fractions approach nearly full ionisation ($[80-90]\%$) only at $r \sim 0.5 R_\mathrm{500c}$ and are especially low ($\lesssim 20\%$) in a shell outside the region immediately heated by AGN feedback.
    \item When interpreting $M_{12}$ as a dynamical state indicator, it bears a moderate correlation ($|\rho^\mathrm{sp}| \sim 0.4$) with electron number density profiles: More disturbed protoclusters have a lower central density and a higher density at intermediate radii. The integrated Eddington ratio $\bar f_\mathrm{Edd}$ correlates especially with central density, accounting for over a factor of four between the highest and lowest bins in $\bar f_\mathrm{Edd}$.
    \item The ANOVA analysis revealed that these three potentially observable quantities ($M_\mathrm{500c}$, $M_{12}$ and $\bar f_\mathrm{Edd}$) provide complementary constraints on the profiles, suggesting that it should be feasible to build a model for $n_e(r)$ that takes the three variables into account.
    \item The electron number densities derived here from \textsc{Magneticum} are systematically lower than the ones derived for the Spiderweb protocluster by \citet{Tozzi_2022} and \citet{Lepore_2024}. 
    While several systematic effects make it difficult to assess the statistical significance of such a discrepancy, it may indicate either limitations of the AGN feedback modelling (beyond the purely thermal energy injection) and/or the possible rarity of the Spiderweb system.
\end{enumerate}

In future work, we will build on these results to present a parametric description of these profiles, and we will calibrate fitting formulae for the electron density profiles in terms of mass, dynamical and central states, and their redshift evolution beyond $z \simeq 2$. More broadly, future work will also need to explore in how far different simulation models (especially regarding their treatment of AGN feedback) converge on the properties of the proto-ICM in order to assess the model dependence of the present results.

\begin{acknowledgements}
We thank the anonymous referee for their insightful comments, which have been useful to improve this work. We also thank S. Ettori, V. Ghirardini, and M. Roncarelli for useful discussions, and A. Ragagnin for support with the access to \textsc{Magneticum} data during the early stages of this project. 
DVP, AB and MB acknowledge support from the ERC CoG $\vec{B}$ELOVED, GA n. 101169773.
KD acknowledges support by the COMPLEX project from the European Research Council (ERC) under the European Union’s Horizon 2020 research and innovation program grant agreement ERC-2019-AdG 882679.
MB acknowledges financial contribution from the INAF GO grant 1.05.24.02.10 {\it Extended Radio Emission in Galaxy Clusters at deep focus with MeerKAT}.
The calculations for the hydrodynamical simulations were carried out at the Leibniz Supercomputer Center (LRZ) under the project pr83li (Magneticum). 
\end{acknowledgements}
\bibliographystyle{aa} 
\bibliography{aa60465-26}

\clearpage
\FloatBarrier

\begin{appendix}

\section{Matching of haloes between \textsc{Subfind} and \textsc{ASOHF}}
\label{s:app.matching_asohf}
For the analyses presented in this work we reprocessed all 3836 extracted regions, which already contained a \textsc{Subfind} halo catalogue, with \textsc{ASOHF} \citep{Planelles_2010, Valles-Perez_2022}. The necessity to do so stems from an undesired effect, present in $\lesssim 10\%$ of the sample, where the centre of the \textsc{Subfind} group (formally, the position of the most bound particle; \citealp{Springel_2001, Dolag_2009}) may closely correspond to the most prominent density peak, but this does not coincide with the centre of spherical symmetry of the object at larger scales.

\begin{figure}
    \centering
    \includegraphics[width=0.9\linewidth]{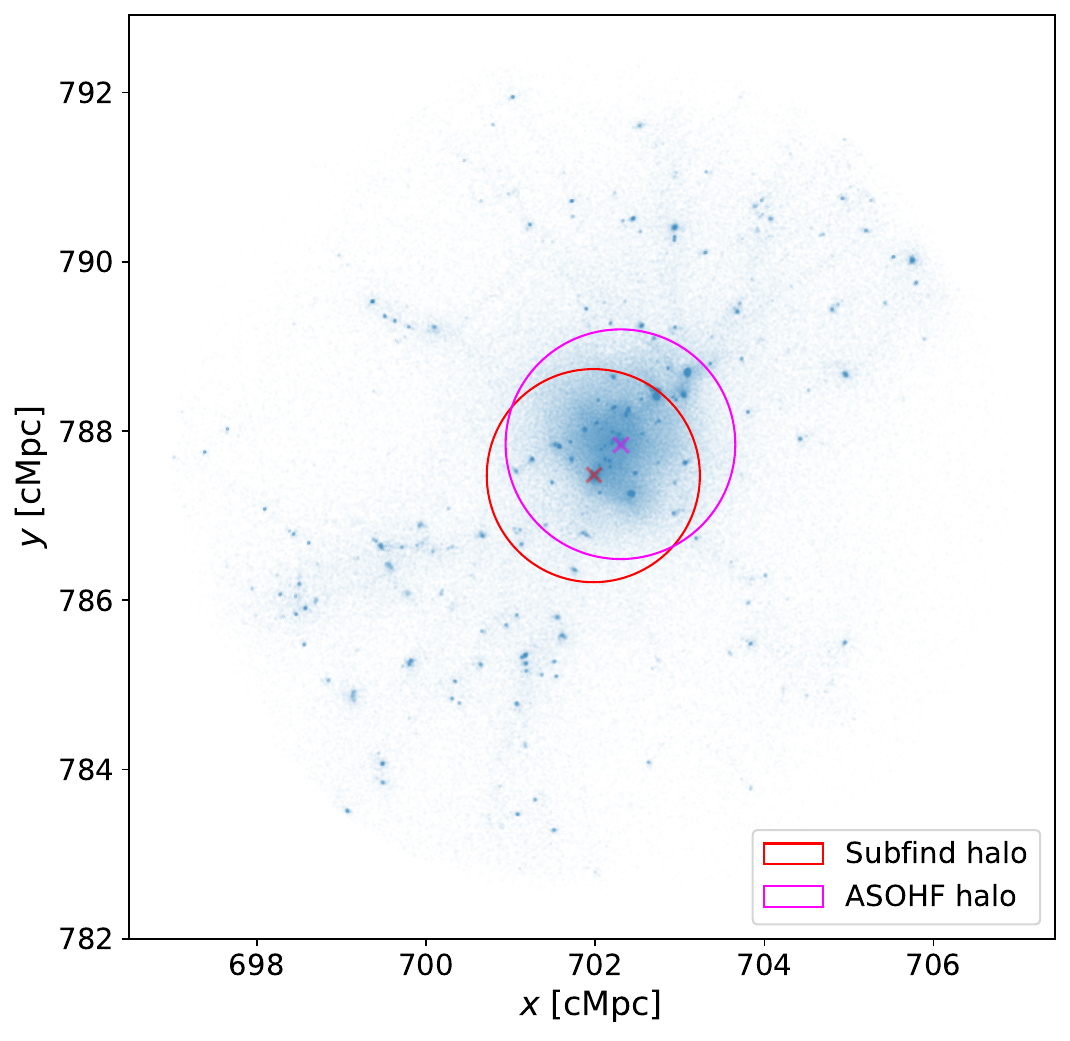}
    \caption{Projection of the gas particle distribution around a massive protocluster (blue dots), with the \textsc{Subfind} halo centre and $R_\mathrm{500c}$ boundary indicated as a red cross and circle, respectively. The best-matching \textsc{ASOHF} halo is likewise indicated in magenta.}
    \label{fig:subfind_fail_map}
\end{figure}

An example of such effect is shown in Fig.~\ref{fig:subfind_fail_map}, where the gas particle distribution around a massive protocluster is shown by the blue dots, and \textsc{Subfind}'s centre and $R_\mathrm{500c}$ are represented as a red cross and circle, respectively. While \textsc{Subfind} centres on the visually strongest density peak, the smooth distribution of particles appears clearly displaced towards the positive $x$ and $y$ directions.

This justifies the necessity to recentre the protoclusters to extract the profiles, which we achieved by running the spherical overdensity halo finder \textsc{ASOHF} on each of the extracted regions. In particular, the best-matching \textsc{ASOHF} halo for a given \textsc{Subfind} region, with radius $R_\mathrm{500c}^\mathrm{Subfind}$ is chosen as the one minimising
\begin{equation}
    \mathrm{ID_{ASOHF}} = \mathrm{argmin}_i \frac{d_i}{R_\mathrm{500c}^\mathrm{Subfind}},
\end{equation}
\noindent where $d_i \equiv |\mathbf{r_\mathrm{Subfind}}-\mathbf{r}_{\mathrm{ASOHF},i}|$, and only ASOHF haloes with $0.5 < M_{\mathrm{ASOHF},i} / M_\mathrm{500c}^\mathrm{Subfind} < 2$ are considered as candidates. The matching also avoids duplicate associations, since a small number of the regions extracted from \textsc{Subfind} do overlap.

\begin{figure}
    \centering
    \includegraphics[width=0.9\linewidth]{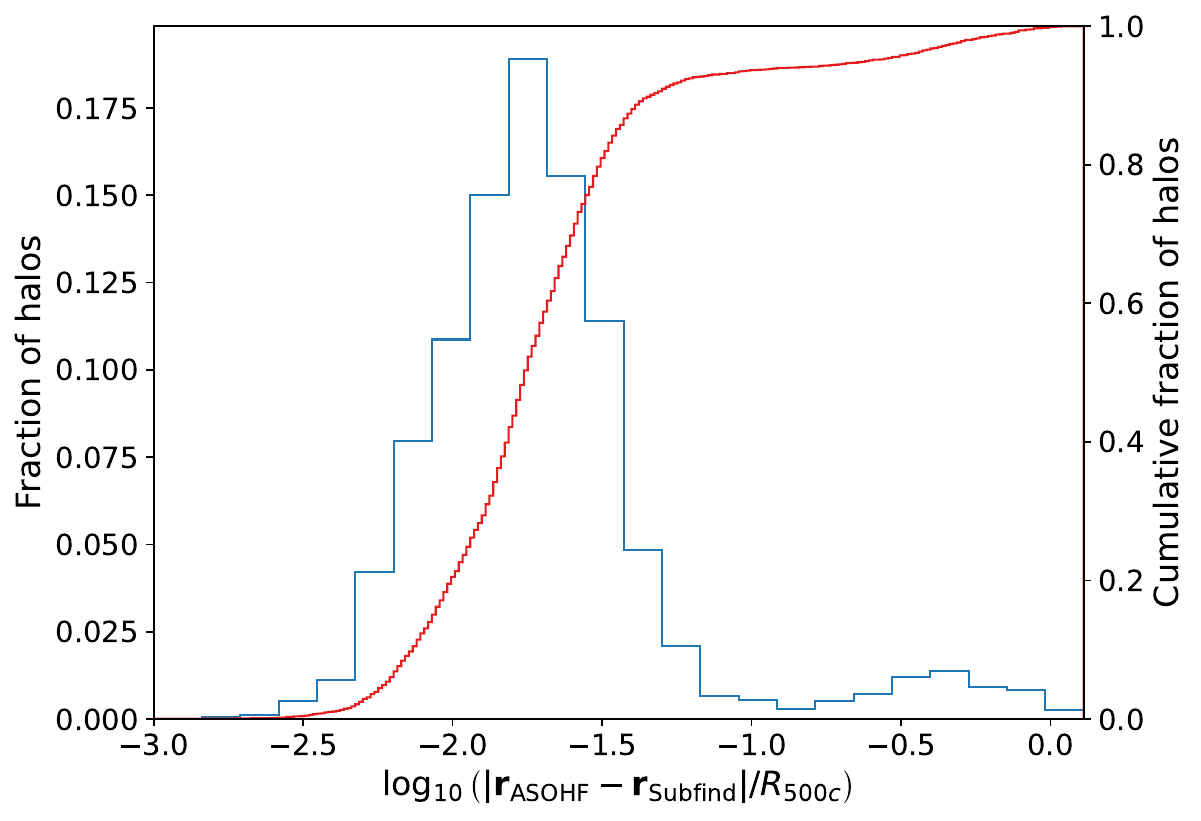}
    \caption{Distribution of offsets between the centre of the \textsc{Subfind} halo, used to extract the $8 R_\mathrm{500c}$ spherical regions, and the centre of the best match to an \textsc{ASOHF} halo (blue histogram, left axis). The red line, to be read on the right axis, contains the cumulative version of the same histogram.}
    \label{fig:subfind_fail_matching}
\end{figure}

The results of such a matching are summarised in Fig.~\ref{fig:subfind_fail_matching}, where the blue histogram (according to the left vertical axis) describes the distribution of centre offset between both halo finders, well below $\sim 3\%$ of $R_\mathrm{500c}$ in over half the cases. The red lines, according to the vertical axis on the right, contains the cumulative version of the histogram, showing that only in $\lesssim 5\%$ of the cases the shift exceeded $0.1 R_\mathrm{500c}^\mathrm{Subfind}$. We checked these cases mostly correspond to merging systems, where the definition of the primary centre might be inherently ambiguous. This process left 15 regions for which no viable match is found, which are removed from the sample, together with three additional regions where \textsc{ASOHF} fails to recover all the required properties of the central halo and its galaxy population.

\section{Choice of $M_{12}$ as an indicator of dynamical state}
\label{s:app.M12}

Throughout the manuscript, we have used the stellar mass ratio between the most massive satellite and the BCG, $M_{12}$, as an indicator of dynamical state. This is grounded on the expectation that, after a major merger, the most massive satellite in the system is likely to be the BCG of the infalling group. Thus, the presence of a very massive satellite may point at a recent merger. While this is well-grounded in the low-redshift literature \citep{Jones_2003, Hearin_2013, Ragagnin_2019, Kimmig_2025}, it could be reasonable to ask if this is still the case at higher $z$, when systems are more disturbed and the BCG might not be univocally defined.

\begin{figure*}
    \centering
    \includegraphics[width=0.8\linewidth]{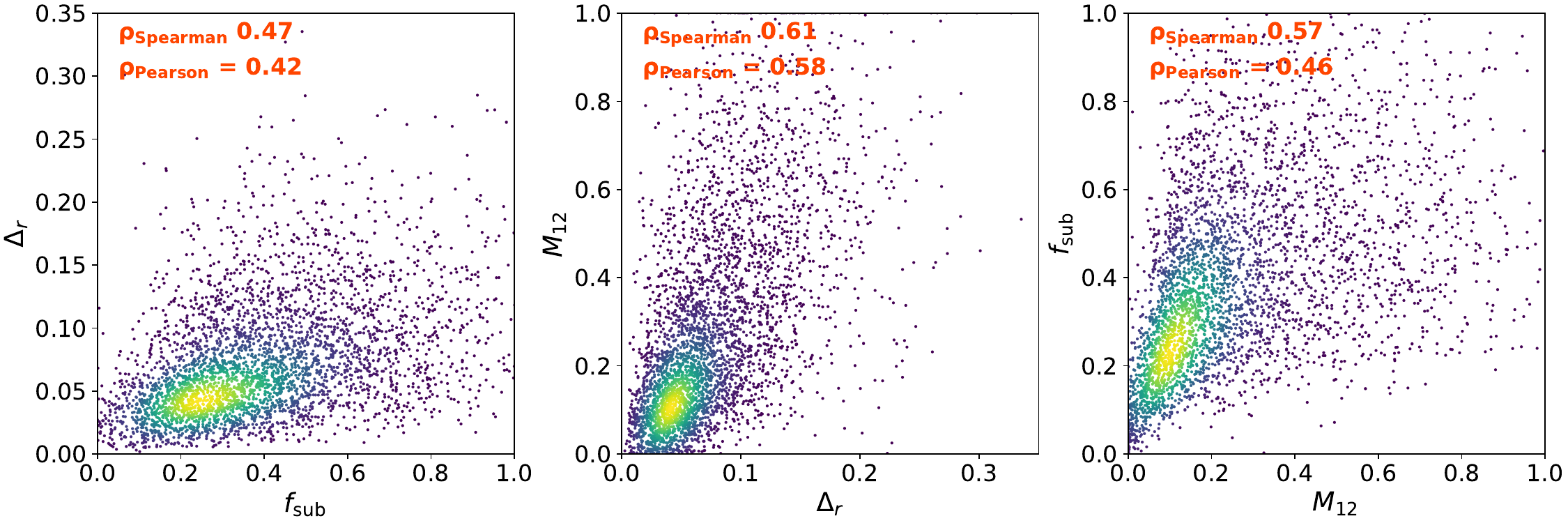}
    \caption{Scatter plots of the pairs of assembly state indicators $f_\mathrm{sub}$ and $\Delta_r$ (left), $\Delta_r$ and $M_{12}$ (middle), and $M_{12}$ and $f_\mathrm{sub}$ (right) computed over the whole sample. In the plots, the colour represents a kernel density estimate of the local point density, to better highlight the trends where the point density is high.}
    \label{fig:dynstate}
\end{figure*}

While a detailed study on the suitability of different indicators of assembly state falls beyond the scope of this work, in Fig.~\ref{fig:dynstate} we partly address this issue by showing the correlations between $M_{12}$ and two widely used indicators of dynamical state in the galaxy clusters literature, namely, the offset between density peak and centre-of-mass, $\Delta_r$, in units of $R_\mathrm{vir}$ \citep{Crone_1996} and the fraction of mass in substructures, $f_\mathrm{sub}$, often used in combination \citep{Neto_2007}. These quantites were computed from the outputs of \textsc{ASOHF}, and we refer the reader to \citet{Valles-Perez_2023} for further details on their calculation.

We observe moderate quasi-linear correlations between any pairs of these indicators ($f_\mathrm{sub}$ and $\Delta_r$, left; $\Delta_r$ and $M_{12}$, middle; and $M_{12}$ and $f_\mathrm{sub}$, right). These correlations, even when measured at $z \simeq 0$, have been reported in the literature to be moderately low \citep[e.g.][]{Haggar_2024, Valles-Perez_2025}. In our case, the magnitude of the Spearman rank-correlation coefficients between $M_{12}$ and the other two quantities is $\rho_\mathrm{sp} \sim 0.6$, slightly higher than what usually found at low $z$. This suggests that, despite the challenges associated to the non-univocality of the BCG, $M_{12}$ remains as a suitable indicator of dynamical state at $z \simeq 2$. 

The choice of this quantity, over the more widely used $\Delta_r$ and $f_\mathrm{sub}$ (especially in the simulations literature), is motivated by its observational practicality. $M_{12}$ can be directly estimated from the stellar masses of the two most massive galaxies within an aperture, which is feasible, although with uncertainties, even at $z \simeq 2$. In contrast, $\Delta_r$ would require very deep observations of the ICM to properly compute a centroid over scales of $R_\mathrm{500c}$ or $R_\mathrm{vir}$, which, as we discuss in this manuscript, is highly challenging. Similarly, $f_\mathrm{sub}$ relies on the identification and mass measurements of multiple substructures. At high-$z$, where angular sizes are small and completeness may be a more significant issue, this would be significantly more complex. This makes $M_{12}$ a better-suited proxy for the purposes of this work.

\section{Further analyses on the density and temperature structures}
\label{s:app.density_mass_additional}

Here we complement the general discussion on density profiles and temperature structure of the proto-ICM (Sect.~\ref{s:results.density_mass}) with further analyses. In Sect.~\ref{s:app.density_mass_additional.temperature} we discuss the gas density fractions in temperature bins (instead of the density fraction above temperature thresholds, as done in Fig.~\ref{fig:gas_density_fractions_T} of the main text). In Sect.~\ref{s:app.density_mass_additional.density} we complement the information about density profiles and baryon fractions in Sect.~\ref{s:results.density_mass}.

\subsection{Further analyses on the temperature structure}
\label{s:app.density_mass_additional.temperature}

\begin{figure*}
    \centering
    \includegraphics[width=0.33\textwidth]{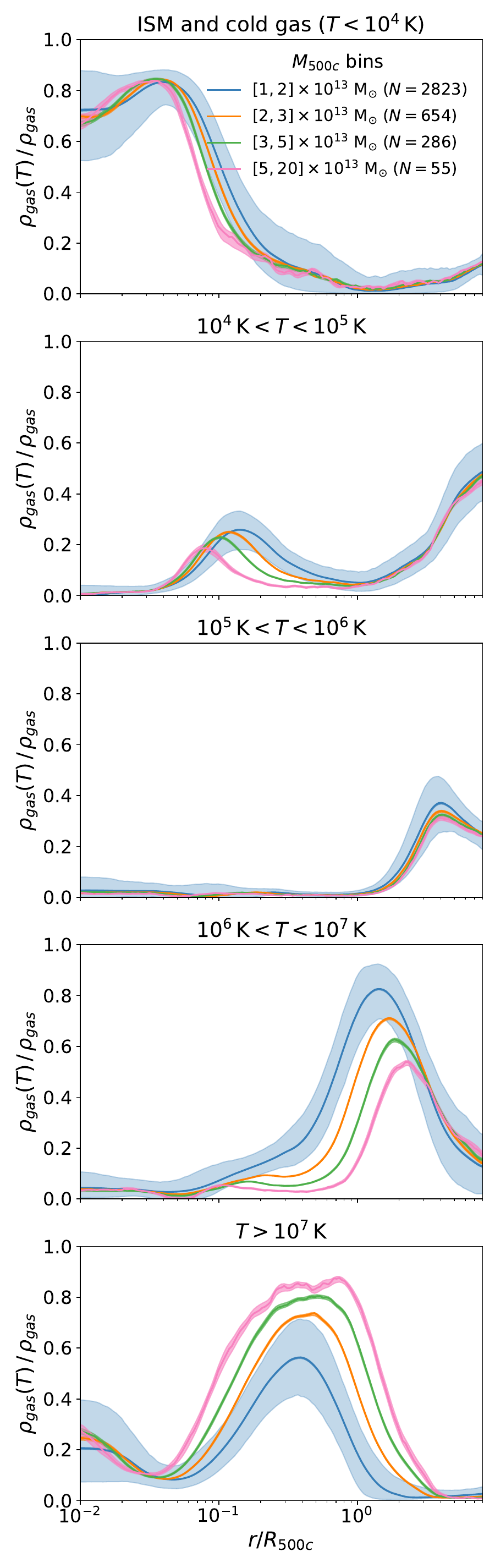}~
    \includegraphics[width=0.33\textwidth]{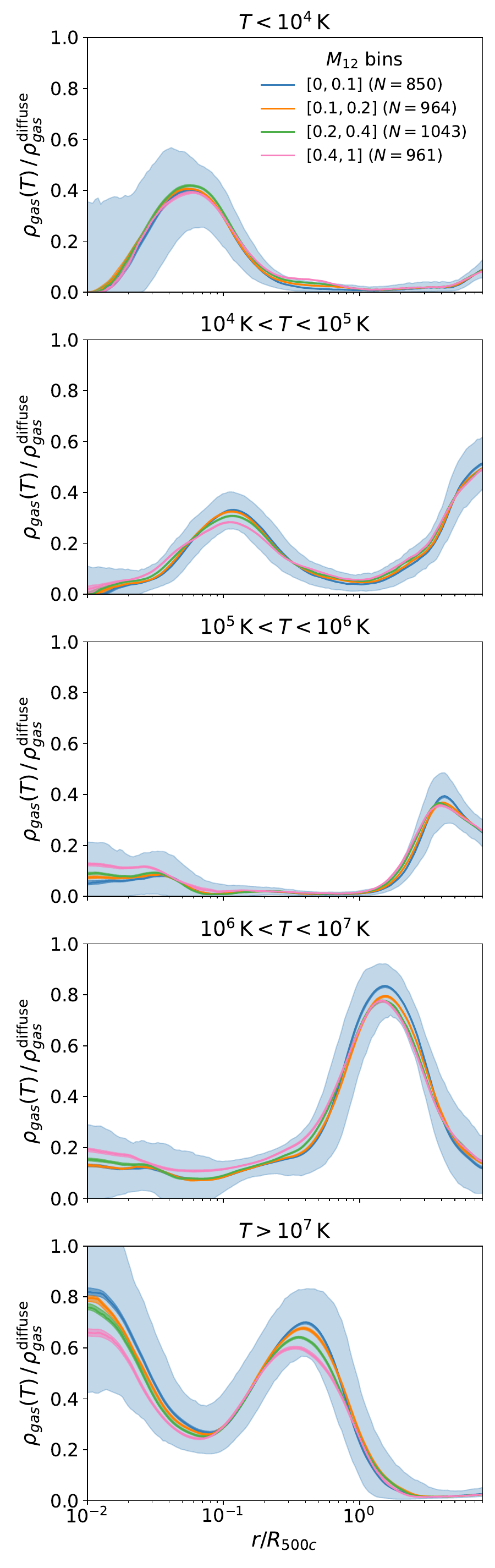}~
    \includegraphics[width=0.33\textwidth]{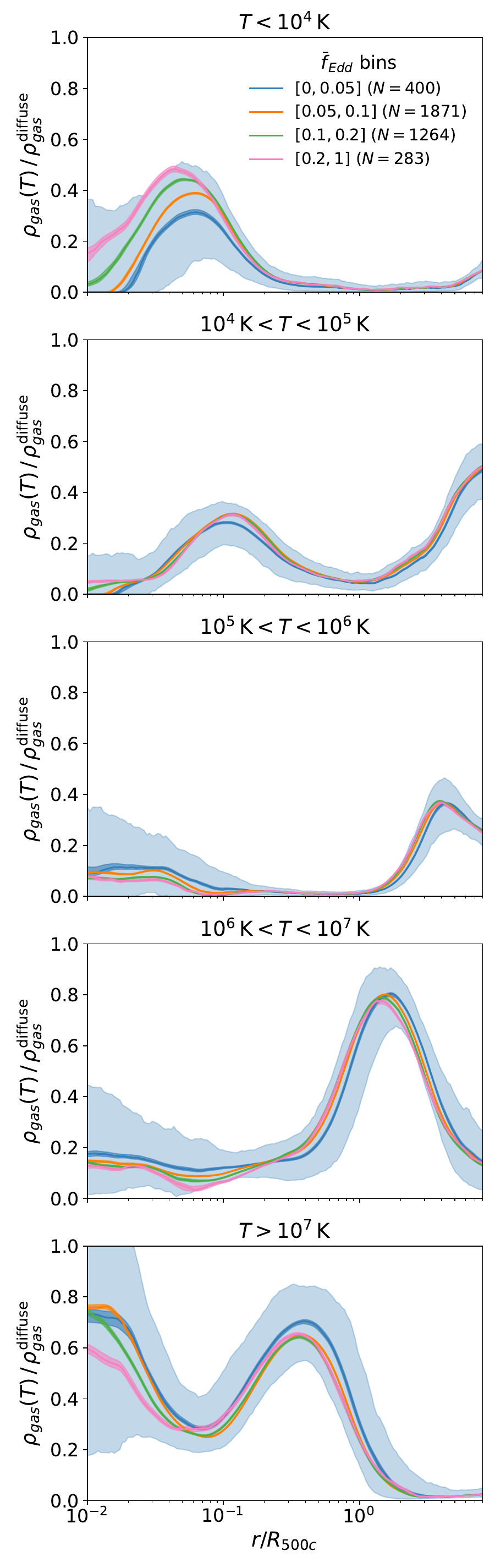}
    \caption{Density fractions in several temperature bins. The left, central and right colums are equivalent to Fig.~\ref{fig:gas_density_fractions_T}, left column of Fig.~\ref{fig:gasfractions_and_correlationsassembly} and middle column of Fig.~\ref{fig:gasfractions_and_correlationsassembly}, respectively, but showing the fractions of mass in several temperature bins, instead of the gas fractions above temperature thresholds.}
    \label{fig:densityfractions_differential}
\end{figure*}

In Fig.~\ref{fig:densityfractions_differential}, we complement the information in Figs.~\ref{fig:gas_density_fractions_T} and \ref{fig:gasfractions_and_correlationsassembly} by showing the profiles of gas mass fractions at several temperature bins (instead of temperatures over a threshold, shown in the main text). In particular, the left column of Fig.~\ref{fig:densityfractions_differential} corresponds to the dependence of the mass fractions on mass (equivalent to Fig.~\ref{fig:gas_density_fractions_T}), while the central and right column correspond, respectively, to the left and middle columns of Fig.~\ref{fig:gasfractions_and_correlationsassembly}, i.e. to the dependence of the mass fractions on $M_{12}$ and $\bar f_\mathrm{Edd}$. While the results shown in the text served us to discuss the abundance of ionised and X-ray emitting gas, these figures can complement the former by showing where the different phases are located.

From the left panel, we observe that ISM and cold ($T < 10^4 \, \mathrm{K}$) gas is the dominant ($\gtrsim 50 \%$) phase within $0.1 R_\mathrm{500c}$, and that this radius does not scale self-similarly, as also found in Sect.~\ref{s:results.density_mass}. Gas in the interval $10^4 \, \mathrm{K} < T < 10^5 \, \mathrm{K}$, which is not fully ionised, builds up in a shell immediately around this region. For massive protoclusters (pink), bremsstrahlung-emitting plasma ($T \gtrsim 10^7 \, \mathrm{K}$) is the dominant phase outside $r / R_\mathrm{500c} \gtrsim 0.1$, while for the lowest-mass protoclusters (blue), these radii are characterised by a mixture of $10^6 \, \mathrm{K} < T < 10^7 \, \mathrm{K}$ and $T >10^7 \, \mathrm{K}$ gas. The outskirts of protocluster haloes, out to $r \sim (3-4)R_\mathrm{500c}$, where the accretion shock is typically located \citep{Molnar_2009, Aung_2021, Valles-Perez_2024}, are dominated by an increasing $10^6 \, \mathrm{K} < T < 10^7 \, \mathrm{K}$ and declining $T >10^7 \, \mathrm{K}$ phases, while outside these radii there is a mixture of cold and hot gas from the diffuse surroundings and smaller haloes.

The dependences of the gas temperature distribution on assembly state, characterised by $M_{12}$, are shown in the middle column, where in concordance with Fig.~\ref{fig:gasfractions_and_correlationsassembly}, we show the fractions with respect to the diffuse gas density (and not to the total gas density; i.e. here we exclude the subgrid ISM from the denominator). Removing the condensed ISM phase, hot gas ($T > 10^7 \, \mathrm{K}$) dominates the centre of the protocluster, but as it becomes more dynamically disturbed ($M_{12}$ increases), its fraction reduces and the contribution gas outside the bremsstrahlung-emitting phase $10^5 \, \mathrm{K} < T < 10^7 \, \mathrm{K}$ increases up to $\sim 30\%$, likely as a result of merger-induced mixing. The fractions of gas in both the $10^6 \, \mathrm{K} < T < 10^7 \, \mathrm{K}$ and $T > 10^7 \, \mathrm{K}$ ranges also correlates negatively with $M_{12}$; generally implying that dynamically disturbed clusters tend to be on average colder (by around $\sim 10\%$ of their mass fraction in different temperature ranges).

Finally, the right panel presents the same fractions, as a function of the Eddington ratio of the central SMBH. Protoclusters with very highly-accreting SMBHs (pink line) tend to host larger fractions of cold and neutral diffuse gas ($T < 10^4 \, \mathrm{K}$) in the central regions ($r \lesssim 0.1 R_\mathrm{500c}$), at the expense of lower fractions of gas in the rest of phases. In the systems with highest accretion rate, gas below $10^4 \, \mathrm{K}$ can reach fractions at $r \sim 0.05 R_\mathrm{500c}$ higher than the bremsstrahlung-emitting gas, thus justifying the pronounced double-$\beta$ structure of the electron number density profiles in Fig.~\ref{fig:electrondensity_othervariables}.

\subsection{Further analyses on the density profiles}
\label{s:app.density_mass_additional.density}

\begin{figure}
    \centering
    \includegraphics[width=0.9\linewidth]{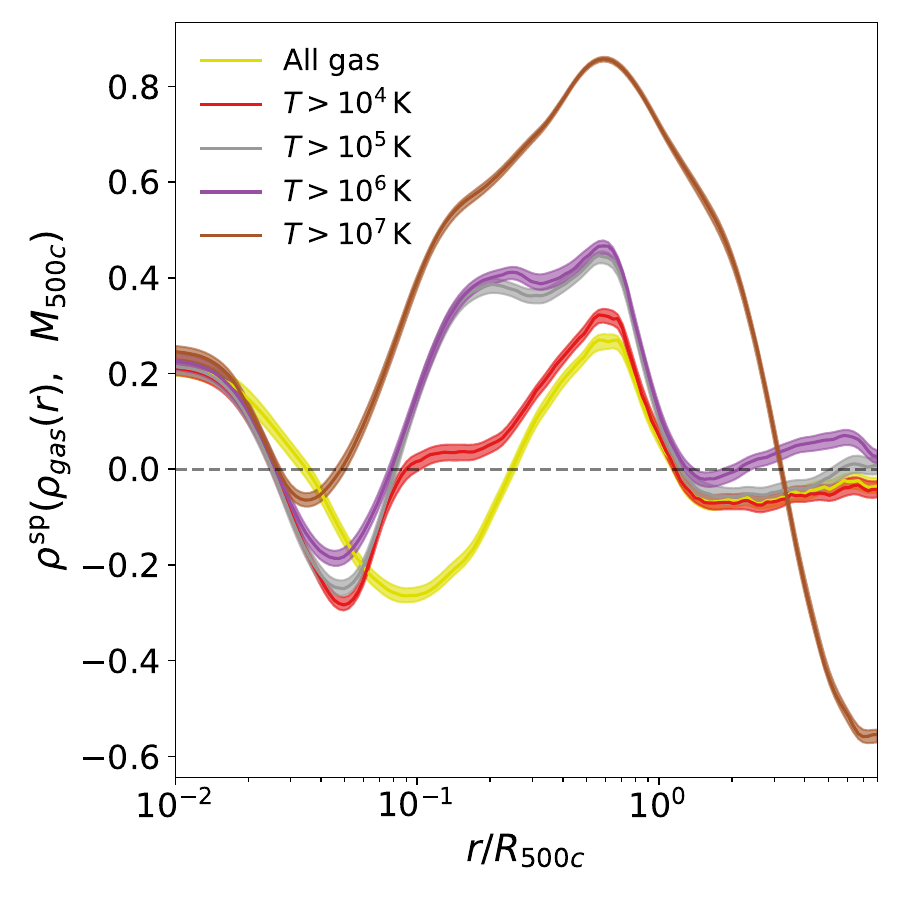}
    \caption{Spearman rank-correlation cofficients between protocluster mass $M_\mathrm{500c}$ and the values of the density profile (for all gas, in yellow, and above several temperature thresholds, in different colours) at each $r/R_\mathrm{500c}$.}
    \label{fig:gas_density_Tthr_correlations}
\end{figure}

The degree to which the trends reported in Fig.~\ref{fig:gas_density_fractions_T} are followed in an object-to-object basis is varied. To discuss it, in Fig.~\ref{fig:gas_density_Tthr_correlations} we show the correlations between the density values at each $r/R_\mathrm{500c}$ and protocluster mass. Gas densities over $10^7 \, \mathrm{K}$ correlate strongly with cluster mass at intermediate radii, as discussed in the main text, due to the self-similar scaling of temperature with mass, $T_{\Delta_c} \propto M_{\Delta_c}^{2/3}$. The departures from self-similarity for the densities of ionised ($\gtrsim 10^5 \, \mathrm{K}$) and hot ($\gtrsim 10^6 \, \mathrm{K}$) gas, on an individual basis, instead, respond to more modest --albeit statistically significant-- correlations of $\rho^\mathrm{sp} \sim 0.4$. Finally, in the central regions there is only a very mild ($\rho^\mathrm{sp} \sim 0.2$) positive correlation of density (at any temperature threshold) with mass. Interestingly, while more massive protoclusters were shown to have slightly lower fractions of cold gas (see Fig.~\ref{fig:gas_density_fractions_T}), the preference of more massive objects towards cuspier density profiles compensates for this effect, and makes this mild correlation independent of the temperature threshold.

\begin{figure}
    \centering
    \includegraphics[width=0.9\linewidth]{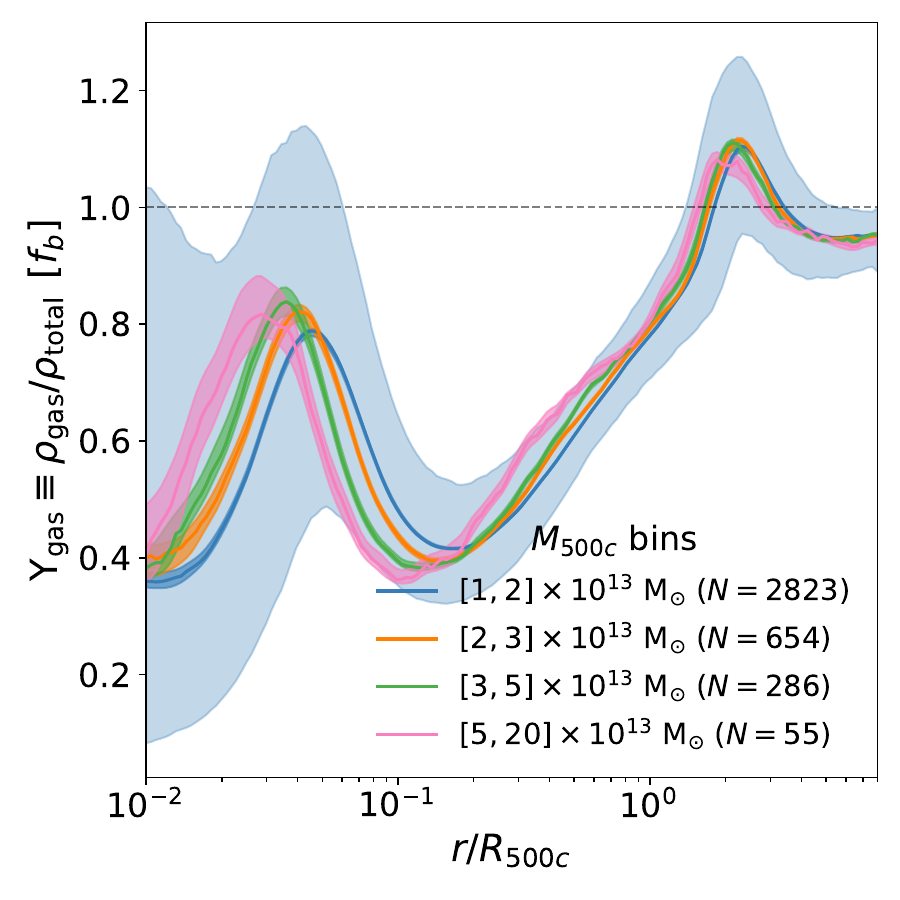}
    \caption{Local gas depletion factors, $\Upsilon_\mathrm{gas}(r) = \rho_\mathrm{gas}(r) / \rho_\mathrm{total}(r)$, stacked by protocluster mass $M_\mathrm{500c}$ according to the legend.}
    \label{fig:gas_depletion_fraction}
\end{figure}

Last, it is also interesting to explore how gas densities compare to DM and stellar densities in protoclusters at $z \simeq 2$. This is presented in Fig.~\ref{fig:gas_depletion_fraction}, where we show the gas depletion function stacked over our mass bins,
\begin{equation}
    \Upsilon_\mathrm{gas}(r) = \frac{\rho_\mathrm{gas}(r)}{\rho_\mathrm{tot}(r)},
\end{equation}

\noindent with $\rho_\mathrm{tot}$ the total (gas, stellar and DM) density. Central $\Upsilon_\mathrm{gas}(r)$ are low due to the steepness of the stellar density profiles, which consistently dominate the central $\sim 50 \, \mathrm{ckpc}$ through our protocluster sample. We then find a local overabundance of gas at $r / R_\mathrm{500c} \in [0.02, 0.05]$ aligned with the location of the dip in hot gas fractions and therefore mostly contributed by cold gas piling up outside the incipient AGN bubbles. Outside $\sim 0.1 R_\mathrm{500c}$, gas fractions consistently increase from $\sim 40\%$ to $\sim 80\%$ of the cosmic value at $\sim 0.8 R_\mathrm{500c}$, with only a small tendency for more massive protoclusters to be relatively more gas rich. Outside the latter radius, we see no mass dependence, and gas fractions continue to increase steeply, and reach the cosmic baryon fraction between $R_\mathrm{500c}$ and $2 R_\mathrm{500c}$.

\section{Estimation for the X-ray exposure times}
\label{s:app.xray_method}

\begin{figure*}
\centering
    \includegraphics[width=0.9\textwidth]{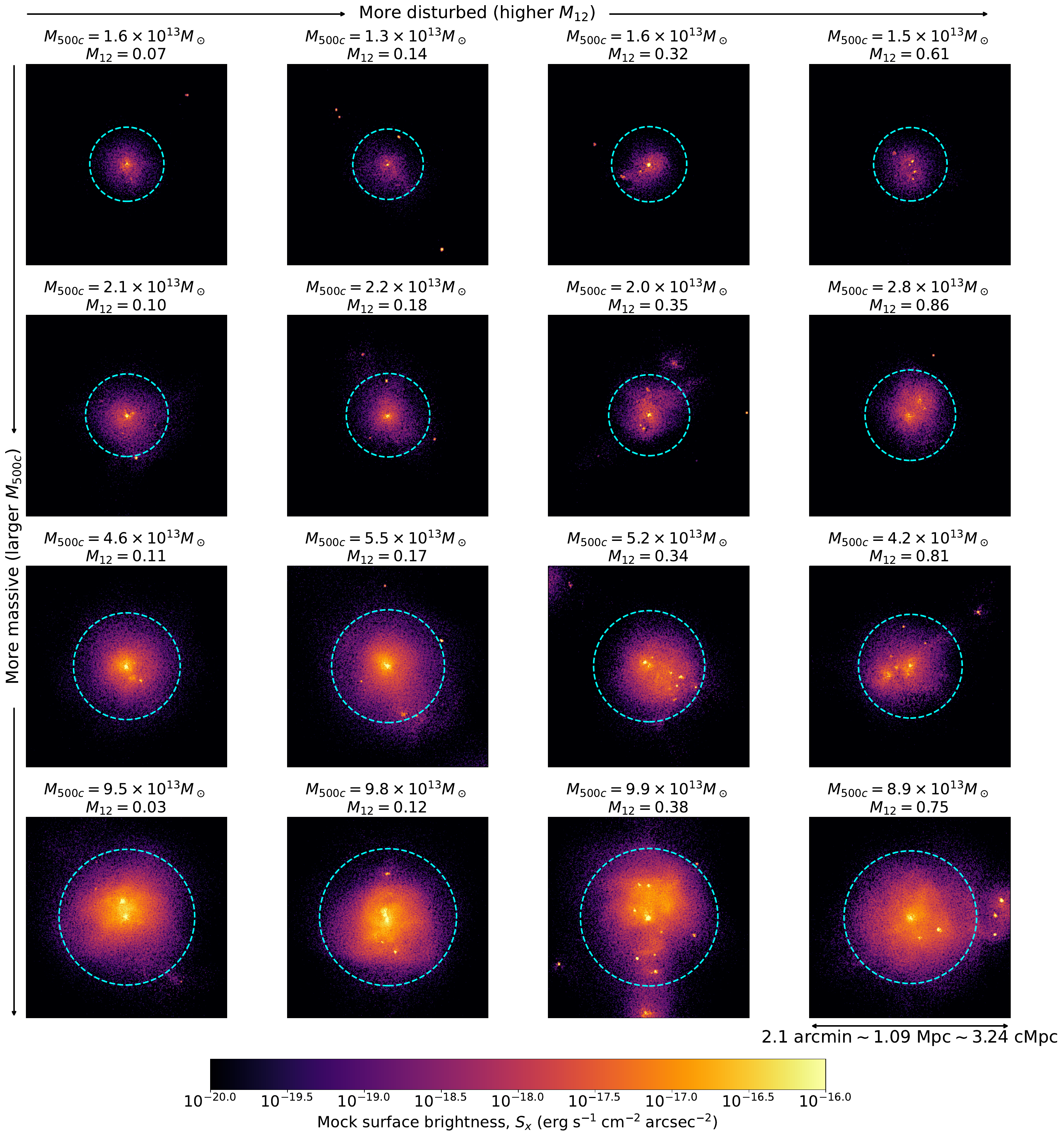}
    \caption{Example mock surface brightness maps for a selection of our simulated protoclusters. Each map corresponds to an angular size of $2.1 \, \mathrm{arcmin}$, corresponding to $1.09 \, \mathrm{Mpc}$ at $z \approx 2$, or $3.24 \, \mathrm{cMpc}$. The blue circle indicates the $R_\mathrm{500c}$ boundary. All maps share the same colourbar. To give a general overview of different masses and dynamical states, mass increases from top to bottom, while the level of dynamical disturbance (as measured by the galaxy mass ratio $M_{12}$) increases from left to right.}
    \label{fig:xraySBmaps}
\end{figure*}

In order to provide a rough estimation of the challenging detectability of the X-ray signal from the proto-ICM in Sect.~\ref{s:discussion.xray}, we produced a simplistic estimation of the X-ray emission from our simulated systems. We computed the X-ray emissivity of each SPH particle in the $[0.5, \, 2] \, \mathrm{keV}$ band (in the observer's frame) assuming an APEC emission model \citep{Smith_2001}. For simplicity, we assumed a solar pattern for the elemental abundances, rescaled by the particle individual metallicities. Gas densities and temperatures are corrected for the cold gas fraction as discussed in Sect.~\ref{s:methods.postprocessing}, and very hot gas particles (which we define\footnote{Other works, such as \citet{Biffi_2025}, use a fix temperature threshold of $5 \times 10^8 \, \mathrm{K}$. Given that we are considering protoclusters, whose physical regime could be fairly different from that of lower-redshift systems, we chose to set a more conservative threshold scaling as the virial temperature, $2 T_\mathrm{500c}$, since it also reasonably selects gas away from the thermal equilibrium of the protocluster due to a recent energy injection. Typically, only $\lesssim 1\%$ of the gas mass corresponds to this phase.} with a temperature threshold of $2 T_\mathrm{500c}$) were excluded from the X-ray emissivity calculations, as it is customarily done in other works using \textsc{Magneticum} \citep[e.g.][]{Biffi_2025}. The motivation for this latter step is related to the AGN feedback modelling in \textsc{Magneticum}, which is only thermal. Consequently, some particles may get spuriously heated to very high temperatures ($\gtrsim 10^8 \, \mathrm{K}$), and still need to thermalise within the ICM.

The APEC spectra (which we interface via the \textsc{PyAtomDB} library; \citealp{Foster_2020}) were then obtained and integrated within the corresponding energy band. We verified this returns emissivities largely in agreement with widely employed fitting formulae for the bremsstrahlung regime at high temperatures \citep[e.g.][]{Rybicki_1979} but also accounts for line emission, which is dominant at the lower temperatures ($k_BT \lesssim [1-3] \, \mathrm{keV}$) that are characteristic of $(1-5) \times 10^{13} \, M_\odot$ systems.

X-ray surface brightness maps were then produced by assuming an arbitrary normal vector, integrating the emissivities and mapping them to a bidimensional grid with resolution of $0.5 \, \mathrm{arcsec}$ (corresponding to $12.7 \, \mathrm{ckpc} \approx 4.7 \, \mathrm{kpc}$ at $z=2$), similar to the on-axis PSF of Chandra \citep{Chandra_2025}. The projected and integrated emissivities are finally converted to flux per pixel by normalising by $4 \pi d_L^2$, $d_L$ being the luminosity distance. In Fig.~\ref{fig:xraySBmaps} we present some example mock surface brightness maps for protocluster of varying mass and dynamical state.

Where necessary, we transform this flux $F_E$ (in units of $\mathrm{erg \, s^{-1} \, cm^{-2}}$) to photon counts by assuming an effective area $A_\mathrm{eff} = 500 \, \mathrm{cm}^2$, exposure time $\Delta t_\mathrm{exp}$ and assuming a representative energy $\langle E_\gamma \rangle = 1 \, \mathrm{keV}$ in the observer's frame,

\begin{equation}
    N_\gamma = \frac{F_E A_\mathrm{eff}\Delta t_\mathrm{exp}}{\langle E_\gamma \rangle}.
\end{equation}

To obtain a simplistic estimate of the required exposure time to detect X-ray emission from the proto-ICM, we first construct radial profiles of count rates, considering eight linearly spaced radial bins from $r=0$ to $r=R_\mathrm{500c}$ (i.e. $\Delta r = R_\mathrm{500c} / 8$). 
We explored masking substructures by removing pixels exceeding by more than $3 \sigma$ the mean logarithmic count rate in each annulus (e.g. \citealp{Zhuravleva_2013}). These high-surface-brightness regions, typically associated with hot gas around galaxies, contribute at the $\sim (5-20)\%$ level to the integrated bin luminosities, but have a very limited impact on the resulting radial trends. That is to say, if substructures are masked in the analysis, the predicted $\Delta t_\mathrm{exp}$ increases by $(5-30)\%$ depending on the mass and radial bin; but the overall trends are not altered. While the dynamics of this gas are fundamentally distinct from the proto-ICM, they still contribute to the X-ray surface brightness. Additionally, masking them in observational data is unfeasible due to their small angular sizes, unlike in low-$z$ systems. For simplicity, and given these considerations, we chose to retain them in the fiducial analysis presented in Sect.~\ref{s:discussion.xray}.
We assume a background rate\footnote{This value is chosen just as an average, representative background level. For instance, \citet{Chandra_2025} quote a typical background count rate of $\sim 0.05 \, \mathrm{cts \, \mathrm{s^{-1}}}$ per chip for ACIS-I, which is a 2x2 array where each chip has a size of $\sim (8 \, \mathrm{arcmin})^2$, for the same energy band we are considering. This corresponds to a background count rate surface density of $\sim 7 \times 10^{-4} \, \mathrm{cts \, s^{-1} \, arcmin^{-2}}$, which is reasonably close to our chosen value (see other similar background levels, e.g. \citealp{Jiang_2008}).} of $b = 5 \times 10^{-4} \, \mathrm{cts \, s^{-1} \, {arcmin}^{-2}}$ and integrate it in the same annuli. This yields the integrated count rates for the signal, $S$, and background, $B$. We finally estimate the required exposure time in order to get a signal-to-noise ratio of $\mathrm{S/N} = 5$ by 

\begin{equation}
    \Delta t_\mathrm{exp} = \mathrm{S/N}^2 \frac{S + B}{S^2},
\end{equation}

\noindent which is what we show in Fig.~\ref{fig:xray}
as a function of protocluster mass and radial bins. The choice of $\mathrm{S/N} = 5$ is, to some extent, arbitrary, but serves to anchor the scaling of our estimation to values consistent with the detection of the diffuse thermal emission from the proto-ICM. It is worth emphasising that considerably higher quality data would still be needed for determining densities through spectral fitting \citep[e.g.][]{Chexmate_2021, Chen_2024}

\section{Further details on the comparison to the Spiderweb protocluster}
\label{s:app.spiderweb}

Within Sect.~\ref{s:discussion.spiderweb}, the lower panel of Fig.~\ref{fig:compare_2_spiderweb} showed our simulated comoving $n_e$ profiles, colour-coded by mass, as a function of comoving radius, together with the profiles derived by \citet{Lepore_2024} as thick, magenta lines. When the comparison is not tied to a potentially uncertain value of $R_{500c}^\mathrm{Spdw}$, we found a better qualitative agreement between the simulated and observed profiles, notwithstanding the fact that, by construction, the latter do not show the inner steepening. In consistence with the analysis in the main text for the averaged densities, simulated $n_e(r)$ with higher $M_{500c}$ than inferred from X-ray and SZ seem to be more consistent with the observed profile, albeit this does not solve the discrepancy as it would imply masses larger than measured by a factor of $\sim 3$.

\begin{figure}
    \centering
    \begin{minipage}[c]{0.35\textwidth}
    \raggedleft
    \includegraphics[width=0.95\textwidth]{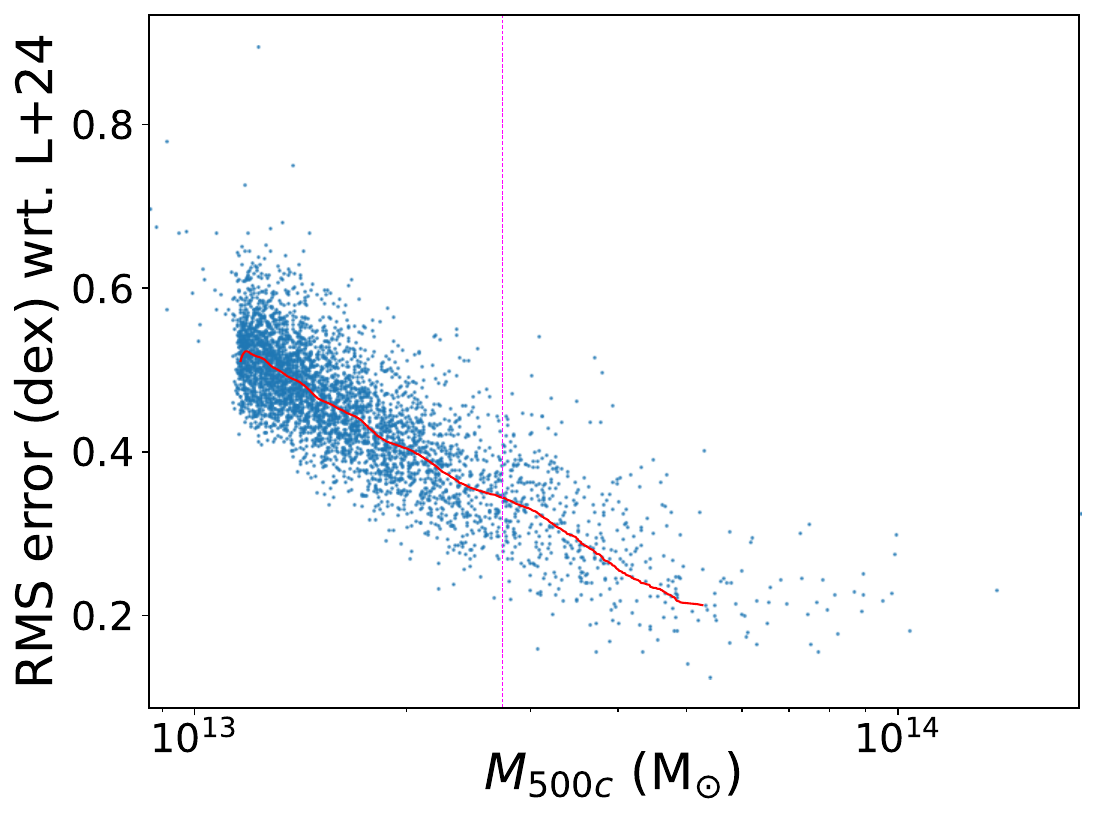}
    \includegraphics[width=\textwidth]{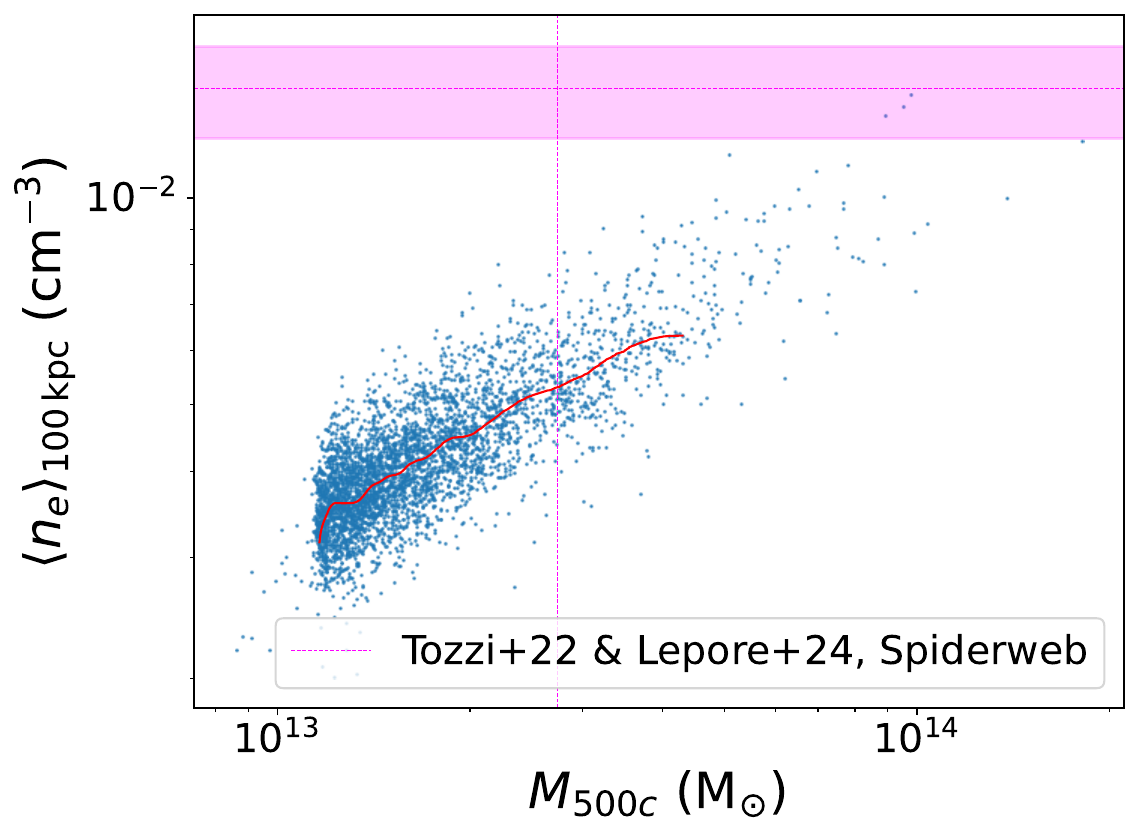}
    \includegraphics[width=\textwidth]{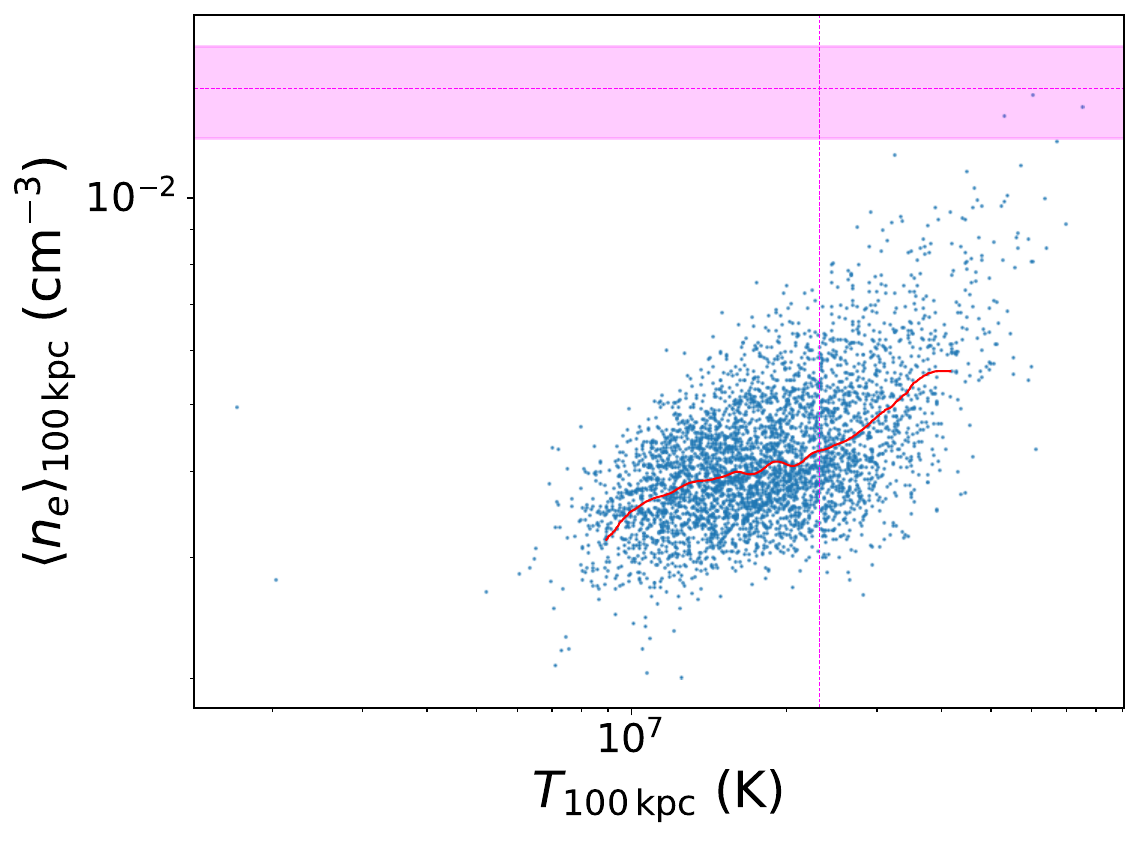}
    \end{minipage}
    \caption{Trends for the match between our simulated $n_e(r)$ profiles and the observational results by \citet{Tozzi_2022} and \citet{Lepore_2024} on the Spiderweb protocluster (magenta). Red solid lines are moving medians to better show the trends. \textit{Top:} RMS error of the profiles with respect to the profile observed by \citet{Lepore_2024}, as a function of protocluster halo mass, $M_\mathrm{500c}$. \textit{Middle:} relation between the mean $n_e$ in the central $100 \, \mathrm{kpc}$ of our simulated protoclusters and $M_\mathrm{500c}$. \textit{Bottom:} relation between the mean $n_e$ in the central $100 \, \mathrm{kpc}$ of our simulated protoclusters and the mass-weighted temperature in the central $100 \, \mathrm{kpc}$, $T_{100 \, \mathrm{kpc}}$.}
    \label{fig:app.spiderweb_further}
\end{figure}

Here we complement the results of Fig.~\ref{fig:compare_2_spiderweb} with some quantitative measurements of the match between our simulated $n_e(r)$ profiles and the one measured by \citet{Lepore_2024} for the Spiderweb protocluster, as a function of the protocluster halo mass. The top panel oof Fig.~\ref{fig:app.spiderweb_further} shows that the root mean square error between observed and simulated profiles,

\begin{equation}
    \mathrm{RMS}_{n_e(r)}= \left[ \frac{\sum_{i=1}^{N_\mathrm{bins}} \left( \log_{10} n_e^\mathrm{Spdw}(r_i) - \log_{10} n_e^\mathrm{sim}(r_i) \right)^2}{N_\mathrm{bins}} \right]^{1/2},
\end{equation}

\noindent decreases sharply with halo mass. Here, $n_e^\mathrm{Spdw}(r_i)$ is the density profile found by \citet{Lepore_2024} with $n_d=4$, and $n_e^\mathrm{sim}(r_i)$ is the density profile of each of our simulated protocluster, interpolated at the same physical radius $r_i$. We complement this with the middle and bottom panels, that show the averaged $n_e$ within $100 \, \mathrm{kpc}$ in our simulated sample ($\langle n_e \rangle_{100 \, \mathrm{kpc}}$) versus $M_\mathrm{500c}$ (middle) and the mass-weighted temperature within 100 kpc ($T_{100 \, \mathrm{kpc}}$, bottom). The values for Spiderweb by \citet{Tozzi_2022} are included for reference as magenta horizontal and vertical lines. This confirms that our simulated $n_e(r)$ seem to match the Spiderweb measurement only for higher values of $M_\mathrm{500c} \gtrsim 8 \times 10^{13} \, M_\odot$, $T_\mathrm{500c} \gtrsim 4 \times 10^{7} \, \mathrm{K}$. Overall, even when not anchoring the comparison to the observationally-determined value of $M_\mathrm{500c}$, only three systems would have central densities within $100 \, \mathrm{kpc}$ as high as the Spiderweb.

\end{appendix}

\end{document}